\documentclass[12pt]{article}
\usepackage{amsmath,amssymb, theorem}
\newtheorem{lemma}{Lemma}
\newtheorem{prop}{Proposition}
\newtheorem{thm}{Theorem}
\newtheorem{coro}{Corollary}

\theorembodyfont{\rm}
\newtheorem{ex}{Example}
\newtheorem{rem}{Remark}

\def\Ae {\mathcal A}
\def\Be {\mathcal B}
\def\Ce {\mathcal C}
\def\Ha {\mathcal H}
\def\Ka {\mathcal K}
\def\Ie {\mathcal I}

\def\Me {\mathcal M}
\def\Ne {\mathcal N}
\def\Oe {\mathcal O}

\def\Se {\mathfrak S}

\def\<{\langle}
\def\>{\rangle}
\def\Tr{\mathrm{Tr}\,}
\def\ptr{\mathrm{Tr}}
\def\supp{\mathrm{supp}\,}

\def\comb{\mathrm{Comb}}

\def\qed{{\hfill $\square$}\vskip 5pt}

\def\<{\langle}
\def\>{\rangle}

\title{Generalized channels: channels for convex subsets of the state space}
\author{Anna Jen\v cov\'a\\
{\small Mathematical Institute, Slovak Academy of Sciences,}\\
{\small \v {S}tef\'{a}nikova 49, 814 73 Bratislava, Slovakia} \\
{\small  jenca@mat.savba.sk}
}
\date{}
\begin{document}

\maketitle

{\textbf{Abstract.} Let $K$ be a convex subset of the state space of a finite dimensional $C^*$-algebra. We study the properties of channels on $K$, which
are defined as affine maps from $K$ into the state space of another algebra,
extending to completely positive maps on the subspace generated by $K$.
We show that each such map is the restriction of a completely positive map on the whole algebra, called a generalized channel. We characterize the set of
generalized channels and also the equivalence classes of generalized channels
having the same value on $K$. Moreover, if $K$ contains the tracial state,
the set of generalized channels forms again a
 convex subset of a multipartite state space, this leads to a  definition of a generalized supermap, which is a generalized channel with respect to this subset.
  We prove a decomposition theorem for generalized supermaps and describe the equivalence classes. The set of generalized supermaps having the same value on equivalent generalized channels is also characterized. Special cases
include quantum combs and process POVMs.
}

\section{Introduction}

The first motivation for this paper comes from the problem of measurement
of a quantum channel. A mathematical framework for such
measurements, or more generally, for measurements on quantum networks,
 was introduced in \cite{daria_circ}, in terms of testers \cite{daria_testers}.  
For quantum channels, these were called process POVMs, or PPOVMs in \cite{ziman}.
Similarly to POVMs, a PPOVM is a collection
of positive operators $(F_1,\dots,F_m)$ in the tensor product
of the input and output spaces, but 
summing up to an operator $I_{\Ha_1}\otimes \omega$
 for some state $\omega$ on the input space. The output probabilities of the
corresponding  channel measurement with values in $\{1,\dots,m\}$ are   then
given by
\[
p_i(\mathcal E)=\Tr (M_i X_{\mathcal E}),\qquad i=1,\dots,m 
\]
where $X_{\mathcal E}$ is the Choi matrix of the channel $\mathcal E$.
Via the Choi isomorphism, the set of channels $\Ce(\Ha_0,\Ha_1)$
 can be viewed as
(a multiple of)  an intersection of the set of states in
$B(\Ha_1\otimes \Ha_0)$ with a self-adjoint
vector subspace $J$. This is a  convex set,
and a measurement on channels can be naturally defined as an affine map from
this set   to the set of probability measures on the set of outcomes.

A natural question arising in this context is the following: are all such affine maps given by PPOVMs? And if so, is this correspondence one-to-one?

Further, the concept of a quantum supermap was introduced in
\cite{daria_super}, which is a  map $B(\Ha_1\otimes\Ha_0)\to B(\Ha_3\otimes\Ha_2)$
sending channels  to channels. It was argued that such a map should be linear and
 completely positive.
 But it is clear that it is enough to consider  completely positive maps
$J\to B(\Ha_3\otimes\Ha_2)$
sending channels to channels.
We may then ask whether all such maps extend to a
completely positive map on $B(\Ha_1\otimes\Ha_0)$, and if this extension
 is unique.

Supermaps on supermaps were  defined similarly, these are the so-called
quantum combs, which are used in description of quantum networks,
\cite{daria_circ,daria}. It was proved that all quantum combs can be represented
 by memory channels, which are given by a sequence of channels connected by an ancilla, these form the ''teeth'' of the comb. The theory of quantum combs was
subsequently used for optimal cloning \cite{daria_clone} and learning
\cite{daria_learn} of unitary transformations and measurements \cite{daria_POVM}.
 As it turns out, the set of all $N$-combs forms again (a multiple of) an intersection of the set of multipartite states by a vector subspace.

To deal with these questions in full generality,  we introduce the notion of a channel
on a convex  subset $K$ of the state space, which is an affine map from $K$
into another
state space, extending to a completely positive map on the vector subspace generated by $K$.
In order to include all channels, POVMs and instruments, and other similar objects, we work with finite
dimensional $C^*$-algebras rather than matrix algebras. We show that each such map can be extended to
a completely positive map on the whole algebra, these maps are called generalized channels (with respect to $K$). Further, a measurement on $K$ is defined as an 
affine map from $K$ into the set of probability distributions and it is shown that  each such measurement is given by a (completely) positive map on the whole algebra if and only if 
$K$ is a section of the state space, that is, an
intersection of the set of states by a linear subspace.
This special kind of a generalized channel is called a generalized POVM. 

We describe the equivalence class of generalized
channels restricting to the same channel on $K$.
 Moreover, we show that
if $K$ contains the tracial state,
 the set of generalized channels, via Choi representation, is   again (a multiple of) a section of some state space, so that
 we may apply our results on the set of generalized channels themselves and
repeat the process infinitely. This leads to the definition of a generalized
 supermap.
We show that the quantum combs and testers are particular cases of generalized 
supermaps,
other  examples treated here include channels and measurements
on POVMs and PPOVMs, and supermaps on instruments. We also describe channels on the set of states having the same output probabilities
 for a POVM or a finite number  of POVMs.

The outline of the paper is as follows: After a preliminary section, 
we  consider  extensions of completely positive maps on subspaces of the algebra
and of positive affine functions on $K$. If the subspace is self-adjoint and 
 generated by its positive elements, then a consequence of Arveson's extension theorem shows that any completely positive map can be extended to the whole algebra.
For positive functionals on $K$, we show that these extend to 
positive linear functionals on the whole algebra if and only if $K$ is a section of the state space. 
These results are used in Section
\ref{sec:gchan} for  extension theorems for channels and measurements 
on $K$. We characterize the generalized channels with respect to $K$ and their
 equivalence classes. We show that a generalized channel can be decomposed to a so-called simple
generalized channel and a channel.

In Section \ref{sec:super}, we prove that the set of generalized channels is again a section of a state space and introduce the generalized supermaps. We give a
characterization of generalized supermaps as sections of a multipartite state
 space and show that the quantum combs are a particular case.  We prove a decomposition theorem for the generalized supermaps, similar  to
the realization of quantum  combs by  memory channels proved in \cite{daria}.
In particular, we show that a generalized comb can be decomposed as a simple generalized channel and a comb.
 Finally, we describe the equivalence
classes for generalized supermaps and consider the set of supermaps having the
same value on equivalence classes.

\section{Preliminaries}

Let $\Ae$ be a  finite dimensional $C^*$-algebra. Then $\Ae$ is
isomorphic to a direct sum of matrix algebras, that is, there are finite
dimensional Hilbert spaces $\Ha_1,\dots\Ha_n$, such that
\[
\Ae\equiv \bigoplus_j B(\Ha_j)
\]
Below we always assume that $\Ae$ has this form, so that $\Ae$ is a subalgebra of block-diagonal
elements in the matrix algebra $B(\Ha)$, with $\Ha=\oplus_j \Ha_j$.  The identity in $\Ae$ will be denoted by $I_\Ae$.
We fix a trace $\ptr_\Ae$ on $\Ae$ to be the restriction of the trace in $B(\Ha)$, we omit the subscript
$\Ae$ if no confusion is possible. If $\Ae=B(\Ha)$ is a matrix algebra,
then we write $I_\Ha$ and $\ptr_\Ha$
instead of $I_{B(\Ha)}$ and $\ptr_{B(\Ha)}$. We will sometimes use the notation
$\Ha_A$, $\Ha_B$ etc. for the Hilbert spaces, and $\Ha_{AB}=\Ha_A\otimes\Ha_B$,
$\ptr_A=\ptr_{\Ha_A}$, $I_A=I_{\Ha_A}$.

If $\Be$ is another $C^*$ algebra, then $\ptr^{\Ae\otimes\Be}_\Ae$ will
denote the partial trace on the tensor product $\Ae\otimes\Be$, $\ptr_\Ae (a\otimes b)=\Tr( a)b$. If the input space is clear, we will sometimes
denote the partial trace just by $\ptr_\Ae$. 	

For $a\in \Ae$, we denote by $a^T$ the transpose of $a$. Note that $\ptr^{\Ae\otimes \Be}_\Ae (x^T)=(\ptr^{\Ae\otimes\Be}_\Ae x)^T$ for $x\in \Ae\otimes \Be$.
If $A\subset \Ae$, then $A^T=\{a^T,\ a\in A\}$.

We denote by $\Ae^h$ the set of all self-adjoint elements in $\Ae$, $\Ae^+$ the
convex cone  of positive elements in $\Ae$ and $\Se(\Ae)$ the set of states on
$\Ae$, which will be identified with the set of density operators in $\Ae$,
that is, elements  $\rho\in \Ae^+$ with $\Tr\rho=1$. If $\rho\in \Se(\Ae)$ is
invertible, then we say that $\rho$ is a faithful state. The projection onto the support
of $\rho$ will be denoted by $\supp(\rho)$.
If $\Ae=B(\Ha)$, then we denote
the set of states by $\Se(\Ha)$. Let $\tau_\Ae$ denote the
tracial state $t_\Ae^{-1}I_\Ae$, here $t_\Ae=\Tr (I_\Ae)$. Later on, we will
need also  the set $\Se_c(\Ae)=\{ a\in \Ae^+, \Tr (c a)=1\}$ for a positive
invertible element $c\in \Ae$, note that   $\Se_{I_\Ae}(\Ae)=\Se(\Ae)$.

The trace defines an inner product in $\Ae$ by $\<a,b\>=\Tr (a^*b)$, with this $\Ae$ becomes a Hilbert space.
If $A\subset \Ae$ is any subset,
then $A^\perp$ will denote the orthogonal complement of $A$. Then $A^{\perp\perp}=:[A]$ is the linear subspace,
spanned by $A$. The subspace spanned by a single element $a$ will be denoted by $[a]$.

Let now $L\subseteq \Ae$ be a (complex) linear subspace.
We denote by  $L^h$ the set of self-adjoint elements in $L$, then $L^h$ is a
real vector subspace in $\Ae^h$.
The subspace  $L$ is self-adjoint if
$a^*\in L$ whenever $a\in L$. In this case  $L=L^h\oplus iL^h$. If also $I_\Ae\in L$, then
$L$ is called an operator system \cite{paulsen}. If $L$ is generated by positive elements, then we say that $L$ is positively generated. If $L_1$ and $L_2$ are subspaces in $\Ae$,
then $L_1\vee L_2$ denotes the smallest subspace containing both $L_1$ and $L_2$, and $L_1\wedge L_2=L_1\cap L_2$.

\subsection{Channels, instruments and POVMs}

Let $\Ha$, $\Ka$ be finite dimensional Hilbert spaces.
For any linear map
$T:B(\Ha)\to B(\Ka)$, there is  an element $X_T\in B(\Ka\otimes \Ha)$,
given by
\begin{equation}\label{eq:choi_matrix}
X_T:= (T\otimes id_\Ha)(\Psi_\Ha),\qquad \Psi_\Ha=\sum_{i,j}|i\>\<j|\otimes |i\>\<j|
\end{equation}
for $|i\>$ a canonical basis in $\Ha$.
Conversely, each operator  $X$ in  $B(\Ka \otimes \Ha)$ defines a
  linear map $T_X:B(\Ha)\to B(\Ka)$ by
\begin{equation}\label{eq:choi_map}
T_X(a)= \ptr_{\Ha}[(I_\Ka\otimes a^T)X] , \qquad a\in B(\Ha)
\end{equation}
It is easy to see that $T_{X_T}=T$ and $X_{T_X}=X$ so that the two maps are each other's inverses.
The matrix $X_T$ is called the Choi  matrix of $T$. We have
\begin{enumerate}
\item[(i)]  $T$ is completely positive (cp) if and only if $X_T\ge 0$, 
\cite{choi}.
\item[(ii)] $T$ is trace-preserving if and only if $\ptr_{\Ka} X_T=I_\Ha$.
\end{enumerate}

Let now $\Ae=\oplus_i B(\Ha_i)$ and $\Be=\oplus_j B(\Ka_j)$ be finite dimensional $C^*$ algebras. For any linear map
$T:\Ae\to \Be$ there are linear maps  $T_{ij}:B(\Ha_i)\to B(\Ka_j)$   such that $T(a_i)=
\oplus_j T_{ij}(a_i)$, $a_i\in B(\Ha_i)$.
It is clear that $T$ is a cp map if and only if all $T_{ij}$ are cp maps. Put
\begin{equation}
X_T:=\oplus_{i,j} X_{T_{ij}}\in \Be\otimes\Ae
\end{equation}
Then it  is easy to see that  equation
(\ref{eq:choi_map}) and both (i) and (ii) hold with $\Ha=\oplus_i\Ha_i$ and $\Ka=\oplus_j \Ka_j$ (hence we may replace $\ptr_\Ha$ and $\ptr_\Ka$ by $\ptr_\Ae$ and
$\ptr_\Be$, similarly for $I_\Ha$ and $I_\Ka$).
The matrix $X_T$ is again called the Choi matrix of $T$.

Next we describe instruments and POVMs as special kinds of channels.
Let
$\Ka_j\equiv \Ka$ for all $j=1,\dots,m$, so that $\Be=\mathbb C^m\otimes B(\Ka)$.
Then  a channel $T:\Ae\to \Be$ is called an instrument
$\Ae\to B(\Ka)$, with values in $\{1,\dots,m\}$, \cite{holevo}. Note that $T$ is a channel if and only if $T_{ij}$ are cp maps, such that
for each $i$, $T_i:=\sum_j T_{ij}$ is a channel $B(\Ha_i)\to B(\Ka)$.  The Choi matrix of
an instrument has the form $X_T=\oplus_i\sum_{j=1}^m |j\>\<j|\otimes X_{ij}$, with \[
\ptr_\Be X_T=\oplus_i\sum_j\ptr_{\Ka} X_{ij}=\oplus_i I_{\Ha_i}=I_{\Ha}=I_\Ae.
\]
Let us now suppose that  $\Ka=\mathbb C$, then $\Be$ is the commutative
$C^*$-algebra $\Be =\mathbb C^m$. A channel $T:\Ae\to \Be$ maps states onto
 probability distributions, hence it is given by a POVM $M_1,\dots,M_m\in \Ae^+$,
$\sum_k M_k=I_\Ae$  as
\begin{equation}\label{eq:POVM}
T(a)=(\Tr M_1 a,\dots,\Tr M_m a)
\end{equation}
The Choi matrix is $X_T=\sum_k |k\>\<k|\otimes M^T_k$, with $\ptr_\Be X_T=\sum_j M^T_j=I_\Ae$.

\subsubsection{The link product}

Let $\Ha_i$ be Hilbert spaces, for $i=1,2,\dots$ and let $\Me\subset \mathbb N$
 be a finite set of indices. We denote $\Ha_\Me:= \bigotimes_{i\in \Me} \Ha_i$.
Let $\Ne\subseteq \mathbb N$ be   another finite set and let $X\in \Ha_\Me$,
$Y\in \Ha_\Ne$ be any operators. The link product of $X$ and $Y$ was defined
in \cite{daria} as  the operator $X*Y\in B(\Ha_{\Me\setminus\Ne}\otimes
\Ha_{\Ne\setminus\Me})$,
 given by
\begin{equation}\label{eq:link}
X*Y=\ptr_{\Me\cap\Ne}[(I_{\Me\setminus\Ne}\otimes Y^{T_{\Me\cap\Ne}})
(X\otimes I_{\Ne\setminus\Me})]
\end{equation}
where $T_{\Me\cap\Ne}$ is the partial transpose on the space $\Ha_{\Me\cap\Ne}$.
In particular, $X*Y=X\otimes Y$ if $\Me\cap \Ne=\emptyset$, and
$X*Y =\Tr (Y^TX)$ if $\Me=\Ne$.

\begin{prop}\label{prop:link}\cite{daria}
The link product has the following properties.
\begin{enumerate}
\item (Associativity) Let $\Me_i$, $i=1,2,3$ be sets of indices, such that
$\Me_1\cap\Me_2\cap\Me_3=\emptyset$. Then for $X_i\in \Ha_{\Me_i}$,
\[
(X_1*X_2)*X_3=X_1*(X_2*X_3)
\]
\item (Commutativity) Let $X\in \Ha_\Me$, $Y\in \Ha_\Ne$, then
\[
Y*X=E(X*Y)E
\]
where $E$ is the unitary swap on $\Ha_{\Me\setminus\Ne}\otimes
\Ha_{\Ne\setminus\Me}$.
\item (Positivity) If $X$ and $Y$ are positive, then $X*Y$ is positive.
\end{enumerate}

\end{prop}

The interpretation of the link product is the following: If $X\in B(\Ha_1\otimes
\Ha_0)$ and $Y\in B(\Ha_2\otimes\Ha_1)$ are the Choi matrices of maps
$T_X:B(\Ha_0)\to B(\Ha_1)$ and $T_Y: B(\Ha_1)\to B(\Ha_2)$, then
$Y*X$ is the Choi matrix of their composition $T_Y\circ T_X$.
For  $X\in B(\Ha_1)$, we have
\begin{equation}\label{eq:link_value}
Y*X=T_Y(X)
\end{equation}

Let now $X\in \Ha_\Me$ be a multipartite operator and
let $\Ie\cup \Oe=\Me$ be a partition
of $\Me$, then $X$ defines a linear map $\Phi_{X;\Ie,\Oe}:\Ha_{\Ie}\to \Ha_{\Oe}$, by
\begin{equation}\label{eq:choi_map_multi}
\Phi_{X;\Ie,\Oe}(a_\Ie)=\ptr_{\Ha_{\Ie}}(I_{\Ha_{\Oe}}\otimes a_\Ie^T)X,\qquad a_\Ie\in
\Ha_{\Ie}
\end{equation}
As it was emphasized in \cite{daria}, $X$ is the Choi matrix of many different
maps, depending on how we choose the input and output spaces $\Ie$ and $\Oe$.
The flexibility of the link product is in that it accounts for these possibilities.
 For example, let $\Me = \Me_1\cup\Me_2\cup\Me_0$ and $\Ne=\Ne_1\cup\Ne_2\cup\Me_0$ be partitions of $\Me$ and $\Ne$. Put $\Phi_X:=
\Phi_{X;\Me_1,\Me_0\cup\Me_2}$ and $\Phi_Y:=\Phi_{Y; \Ne_1\cup\Me_0,\Ne_2}$.
Then $Y*X$ is the Choi matrix of the map $B(\Ha_{\Me_1\cup\Ne_1})\to
B(\Ha_{\Me_2\cup\Ne_2})$, given by
\[
\Phi_{Y*X;\Me_1\cup\Ne_1,\Me_2\cup\Ne_2}=(\Phi_Y\otimes id_{\Me_2})
\circ(id_{\Ne_1}\otimes \Phi_X)
\]

In the case when the input and output spaces are fixed, we will often treat a cp
 map  and its Choi matrix as one and the same object, to shorten the discussion.

\section{Extensions of cp maps and positive functionals}

The main goal of this paper is to study cp maps and channels  from a convex subset $K$ of the state space into 
another $C^*$-algebra. To characterize such maps, it is crucial to know whether or when these can be extended to
cp maps on the whole algebra.
This section contains an extension theorem for cp maps on a vector subspace. We also prove that positive affine functionals on $K$ have positive extensions if and only if $K$ is a section, that is an intersection of the state space by a vector subspace.

\subsection{An extension theorem for cp maps}

Let  $J\subseteq \Ae$ be a subspace and let $\Ka$ be a finite dimensional
Hilbert space. Let $\Be\subseteq B(\Ka)$ be a $C^*$-algebra.

A map $\Xi:J\to \Be$ is positive if it maps $J\cap\Ae^+$ into
 the positive cone $\Be^+$ and $\Xi$ is completely positive if  the map
\[
id_{\Ka_0}\otimes \Xi: B(\Ka_0)\otimes J\to B(\Ka_0)\otimes\Be
\]
is positive, for every finite dimensional Hilbert space $\Ka_0$. If $J$ is an operator system, that is
 a self-adjoint subspace containing the unit, 
 then Arveson's extension theorem \cite{arveson,paulsen} states that  any completely positive map
 $\Xi:J\to B(\Ka)$ can be extended to a cp map $\Ae\to B(\Ka)$.

The following is a consequence  of this  
 theorem in finite dimensions.

 \begin{thm}\label{thm:arveson}  Let $J\subseteq \Ae$ be a self-adjoint positively generated subspace.
  Then any cp map $J\to \Be$ can be
  extended to a cp map $\Ae\to \Be$.
 \end{thm}

{\it Proof.} Let $J^+=J\cap \Ae^+$, so that $J$ is generated by $J^+$. There is some $\rho\in J^+$ such that the support of
$\rho$ contains the supports of
 all other elements in $J^+$. Let us denote $p:=\supp(\rho)$, then $J$ is a subspace in the algebra $\Ae_p:=p\Ae p$.
Denote
\[
\Delta: \Ae_p\to \Ae_p,\qquad \Delta(a)=\rho^{1/2}a\rho^{1/2}
\]
Then $J':=\Delta^{-1}(J)$ is an operator system in $\Ae_p$. Moreover,  $\Xi:J\to \Be$ is a cp map
if and only if $\Xi':= \Xi\circ \Delta$ is a cp map $J'\to \Be\subseteq B(\Ka)$. By Arveson's extension theorem,
 $\Xi'$ can be extended to a cp map $\Phi':\Ae_p\to B(\Ka)$. Let $E_\Be:B(\Ka)\to \Be$ be the trace preserving
 conditional expectation,
 then $\Phi:=E_\Be\circ \Phi'\circ\Delta^{-1}$ is a cp map $\Ae_p\to \Be$  extending $\Xi$. This can be obviously extended to $\Ae$.

\qed

\subsection{Sections of the state space}

Let $f$ be an affine function $\Se(\Ae)\to \mathbb R^+$. Then, since
$\Se(\Ae)$ generates the positive cone $\Ae^+$, $f$ can be extended to a positive linear functional on
 $\Ae$. Below  we discuss the possibility of such extension
if $f$ is defined on some convex subset $K\subset \Se(\Ae)$.
 Let us first describe a special type of such subset.

Let $K\subseteq \Se(\Ae)$ be a convex subset and let $Q$ be the convex cone generated by
$K$, then $Q=\{\lambda K,\lambda\ge 0\}\subseteq \Ae^+$. The vector subspace
$[K]$ generated by $K$ is
self-adjoint and  $[K]=Q-Q+i(Q-Q)$.

We say that $K$ is a section of $\Se(\Ae)$
 if
\begin{equation}\label{eq:section}
K=[K]\cap \Se(\Ae).
\end{equation}
It is clear that a section of $\Se(\Ae)$ is convex and compact. It is also
clear that (\ref{eq:section}) is equivalent with
\begin{equation}\label{eq:section_Q}
Q=[K]\cap\Ae^+
\end{equation}

Sections of the state space can be characterized as follows.

\begin{prop}\label{prop:section_Q} Let $K\subset \Se(\Ae)$ be a compact convex subset and let
$Q=\{\lambda K,\lambda\ge 0\}$. Then $K$ is the section of $\Se(\Ae)$ if and only if
$a,b\in Q$ and $b\le a$ implies $a-b\in Q$.
\end{prop}

{\it Proof.} Since we always
have  $Q\subseteq [K]\cap \Ae^+$, it is enough consider the inclusion
 $[K]\cap \Ae^+\subseteq Q$.
But $[K]\cap \Ae^+=(Q-Q)\cap \Ae^+$ and hence any element
$y\in [K]\cap \Ae^+$ has the form  $y=a-b$ with $a,b\in Q$ and $b\le a$.

\qed

\begin{prop} Let $K\subseteq \Se(\Ae)$. Then $K$ is a section of $\Se(\Ae)$ if and only if there is a subspace $J\subseteq \Ae$, such that
$K=J\cap \Se(\Ae)$.
\end{prop}

{\it Proof.} If $K$ is a section of $\Se(\Ae)$, then we can put
$J=[K]$. Conversely, let $K=J\cap \Se(\Ae)$ for some subspace $J\subseteq \Ae$.
Then $Q=J\cap \Ae^+$ and if $a,b\in Q$ with $b\le a$, then obviously $a-b\in J\cap \Ae^+=Q$. By Proposition \ref{prop:section_Q}, $K$ is a section of $\Se(\Ae)$.

 \qed

Note that if $K=J\cap \Se(\Ae)$ for some subspace $J$, we do not necessarily
have $J=[K]$, even if $J$ is self-adjoint.  The next Proposition
clarifies this situation.

\begin{prop}\label{prop:sec_L} Let $J\subseteq \Ae$ be a self-adjoint subspace and let
$K=J\cap \Se(\Ae)\ne \emptyset$. Then there is a projection $p\in \Ae$, such that
 $[K]= J\cap \Ae_p$. In particular,
$J=[K]$ if $J$ contains a positive invertible element.

\end{prop}

{\it Proof.} Suppose first that $J$ contains a positive  invertible element $\rho$
and let $K=J\cap\Se(\Ae)$, equivalently, $Q=J\cap \Ae^+$. Since $\Ae$ is finite dimensional, for any $a\in J^h$,
there is some $M>0$, such that $a\le M\rho$, and then
\[
a=M\rho-(M\rho-a)\in Q-Q
\]
This implies $J^h=Q-Q$ and since $J$ is self-adjoint, $J=[K]$.

For the general case, 
choose some state $\rho\in K$ such that its support
  contains the supports of all $\sigma\in K$, so that $K\subseteq \Ae_p$, where $p:=\supp(\rho)$.
Then  $J_p:=J\cap \Ae_p$ is a subspace in $\Ae_p$, containing  the
 positive invertible  element $\rho$
 and $K=J_p\cap \Se(\Ae_p)$. Hence, by the first part of the proof,  $[K]=J_p$.

\qed

\subsection{Positive affine functions on $K$}

Let   $A(K)$ be the vector space of  real affine functions and $A(K)^+$ the convex cone of positive affine
functions over $K$. In this paragraph, we study elements in $A(K)^+$ that can be
extended to a positive affine functional on $\Se(\Ae)$, hence are given by  positive elements in $\Ae$.

Any element in $A(K)$ extends to a (unique) real linear functional on $[K]^h$
and conversely,
 any linear functional on $[K]^h$ defines an element in $A(K)$, so that
\[
A(K)\equiv ([K]^h)^*\equiv \Ae^h|_{K^{\perp}}:=\{ a +K^\perp, \ a\in \Ae^h\}
\]
In other words, any element $\phi\in A(K)$ has the form $\phi(\sigma)=\Tr a\sigma$ for some $a\in \Ae^h$ and
 two elements $a_1,a_2\in \Ae^h$ define the same  $\phi\in A(K)$ if and only if $a_1=a_2+x$ for some $x\in K^\perp$.

Let $\pi_{K^\perp}: a \mapsto a+K^\perp$ be the quotient map. Then it is clear that
$\pi_{K^\perp}(\Ae^+)\subseteq A(K)^+$.  We are interested in the converse.
Note that if $\bar K$ is the closure of $K$, then $\bar K$ is convex and
$K^{\perp}=\bar K^{\perp}$, $[K]=[\bar K]$ and $A(K)=A(\bar K)$, $A(K)^+=A(\bar K)^+$.

\begin{thm}\label{thm:ext} Let $K\subseteq \Se(\Ae)$ be a nonempty convex subset. Then
$A(K)^+=\pi_{K^\perp}(\Ae^+)$ if and only if $\bar K$ is a section of $\Se(\Ae)$.
\end{thm}

{\it Proof.} It is clear by the remark preceding the Theorem that we may suppose that $K$ is closed.

Let  $K$ be  a section of $\Se(\Ae)$, then any positive affine function on $K$ extends to a positive linear functional
on $[K]$. Since positive functionals are completely positive and $[K]$ is positively generated, the assertion  follows by
Theorem  \ref{thm:arveson}.

Conversely, suppose that $K$ is not a section of $\Se(\Ae)$. Then there is some $x\in [K]\cap \Ae^+$,
 such that $x\notin Q$. Since $Q$ is closed and convex,  by Hahn-Banach separation theorem there is a
 linear functional
 $f$ on $\Ae^h$, such that $f(x)<s<\inf\{f(a), a\in Q\}$, for some $s\in \mathbb R$.
 This implies that $s<f(0)=0$ and, moreover,
 $\lambda f(\sigma)>s$ for all $\lambda\ge 0$, $\sigma\in K$, hence $f(\sigma)\ge 0$ and $f$ defines
 an element $\phi\in A(K)^+$. But $\phi$ has a unique extension to $[K]$, namely $f$, and $f(x)<s<0$, so that
 $\phi$ cannot be given by an element in $\Ae^+$.

\qed

\section{Generalized channels}\label{sec:gchan}

Let $K\subseteq \Se(\Ae)$ be a convex set and
let $\Xi: K\to \Be^+$ be an affine map. Then $\Xi$ extends to a linear map
$[K]\to \Be$. (Note that in general, this extension does not
 need to be positive.)
We will say that $\Xi$ is a cp map  on $K$ if this extension of $\Xi$ is
completely positive. If $\Xi$ also preserves trace (equivalently, $\Xi(K)\subseteq \Se(\Be)$), then $\Xi$ will be called a channel on $K$.

\begin{rem}\label{rem:gch} Note  that  by this definition,
$\Xi$ is a cp map (resp. channel) on $K$ if and only if
(the extension of) $\Xi$ is  a cp map (resp. channel) on
$\tilde K:=[K]\cap \Se(\Ae)$,
the smallest section of $\Se(\Ae)$ containing $K$. Therefore without any loss
of generality we may  suppose that $K$ is a section of $\Se(\Ae)$.
\end{rem}

\begin{thm}\label{thm:gchannel} Let $K\subseteq \Se(\Ae)$ be a convex subset. Then any cp  map on $K$
 has a cp extension to $\Ae$. If $\Phi:\Ae \to \Be$ is a cp map, then $\Phi$ defines a channel on $K$
  if and only if its Choi matrix satisfies
\begin{equation}\label{eq:generalized_channel}
\ptr_\Be X_\Phi \in I_\Ae+(K^T)^\perp
\end{equation}
Two cp maps $\Phi_1,\Phi_2:\Ae\to \Be$ define the same cp map  on $K$ if and only if
\begin{equation}\label{eq:gchan_equiv}
X_{\Phi_1}-X_{\Phi_2}\in \Be \otimes (K^T)^\perp
\end{equation}
\end{thm}

{\it Proof.} Since $[K]$ is positively generated, the first statement follows from Theorem  \ref{thm:arveson}.
The map  $\Phi$ defines a channel on $K$ if and only if
$\Tr (\Phi(a))=1$ for all $a\in K$, that is
\[
\Tr (a^T)=1=\Tr (\Phi(a))=\Tr ((I_\Be \otimes a^T)X_\Phi)=\Tr (a^T \ptr_\Be X_\Phi),\qquad a\in K,
\]
equivalently,  $\ptr_\Be X_\Phi\in I_\Ae+(K^T)^\perp$. Furthermore,
$\Phi_1$ and $\Phi_2$ have the same value on $K$  if and only
if
\[
\Tr (b(\Phi_1(a)-\Phi_2(a)))=\Tr(b\otimes a^T)(X_{\Phi_1}-X_{\Phi_2})=0,\qquad \forall a\in K,\ b\in \Be
\]
that is, $X_{\Phi_1}-X_{\Phi_2}\in (\Be\otimes K^T)^\perp =
\Be\otimes (K^T)^\perp$.

\qed

Any cp map $\Phi:\Ae\to \Be$, satisfying (\ref{eq:generalized_channel})
 will be called a generalized channel. Two generalized channels
having the same value on $K$  will be called equivalent. If we want to stress
the set $K$ (or the subspace $[K]$), we will say that $\Phi$ is a generalized channel with respect to $K$
 (or $[K]$).

We will next introduce an example that will be used repeatedly throughout the paper.
Let $\Ae_0$ be a finite dimensional
$C^*$ algebra and let $S:\Ae \to \Ae_0$, $T:\Ae_0\to \Ae$ be completely positive maps. Let $J_0\subseteq \Ae_0$
be a self-adjoint vector subspace. Then
$S^{-1}(J_0)=\{a\in \Ae,\ S(a)\in J_0\}$ and $T(J_0)$
are self-adjoint subspaces in $\Ae$.
In particular, if $J_0=[S(\rho)]$ is the one-dimensional subspace generated by $S(\rho)$ for  some $\rho\in \Se(\Ae)$,
then $S^{-1}(J_0)\cap \Se(\Ae)$ is the equivalence class containing $\rho$
for the  equivalence relation on $\Se(\Ae)$ induced by $S$.

\begin{lemma}\label{lemma:perp_0} Let $S:\Ae\to \Ae_0$ be a cp map and let
$J_0$ be a  subspace in $\Ae_0$. Then  $S^{-1}(J_0)^\perp=S^*(J_0^\perp)$,
where $S^*:\Ae_0\to \Ae$ is the adjoint of $S$ with respect to $\<a,b\>=\Tr(a^*b)$.

\end{lemma}

{\it Proof.} Let $a\in \Ae$, then
$\Tr( a^* S^*(b))=\Tr (S(a^*)b)=\Tr (S(a)^*b)=0$ for all $b\in J_0^\perp$ if and only if $S(a)\in J_0$,
this implies that $S^*(J_0^\perp)^\perp=S^{-1}(J_0)$, so that $S^{-1}(J_0)^\perp=S^*(J_0^\perp)$.

\qed

We denote by
$S^T$ the linear map $\Ae\to \Ae_0$, defined by $S^T(a)=[S(a^T)]^T$. Note that
the Choi matrix  of $S^T$ satisfies $X_{S^T}  =X_S^T$, so that $S$ is a channel if and only if $S^T$ is a channel.

\begin{lemma}\label{lemma:S_chan}
 Let $S:\Ae\to \Ae_0$ be  a channel  and let $J_0\subseteq \Ae_0$ be a subspace.
 Let $J=S^{-1}(J_0)$. Then
 \begin{enumerate}
\item[(i)] $(J^T)^\perp=(S^T)^*((J_0^T)^\perp) $
\item[(ii)] $I_\Ae+(J^T)^\perp=(S^T)^*(I_{\Ae_0}+(J_0^T)^\perp)$
 \end{enumerate}

\end{lemma}

{\it Proof.} We have
\[
S^{-1}(J_0)^T=\{a,\ S(a^T)\in J_0\}=\{a,\ S^T(a)\in J_0^T\}=(S^T)^{-1}(J_0^T)
\]
(i) now follows by Lemma \ref{lemma:perp_0} and (ii) follows from the
fact that $S^T$ is a channel, so that $(S^T)^*$ is unital.

\qed

\begin{ex}\textbf{(Channels on channels)}\label{ex:channels} Let $\Ae=\Be_1\otimes\Be_0$, $\Ae_0=\Be_0$ and let
$S:\Be_1\otimes\Be_0\to \Be_0$ be the partial
trace $\ptr_{\Be_1}$. Let $J_0=[I_{\Be_0}]=\mathbb CI_{\Be_0}$.  The set
\begin{equation}\label{eq:channelsB0_B1}
\Ce(\Be_0,\Be_1):=\ptr_{\Be_1}^{-1}([I_{\Be_0}])\cap t_{\Be_0}\Se(\Be_1\otimes\Be_0),
\end{equation}
is the  set of all Choi matrices of channels $\Be_0\to \Be_1$. Denote
$J:= \ptr_{\Be_1}^{-1}([I_{\Be_0}])$ and $K= J\cap\Se(\Be_1\otimes\Be_0)$, then $K$ is a section of the state space and
 $\Ce(\Be_0,\Be_1)=t_{\Be_0}K$. It follows that
 $\Xi$ is a channel on $K$ if and only if $t_{\Be_0}^{-1} \Xi$ is a channel on $\Ce(\Be_0,\Be_1)$.
Hence any channel  $\Ce(\Be_0,\Be_1)\to \Se(\Be)$ is given by a cp map $\Phi: \Be_1\otimes\Be_0\to \Be$,
 such that $\ptr_\Be X_\Phi\in t^{-1}_{\Be_0} I_{\Be_1\otimes\Be_2}+(K^T)^\perp$.

Since  $I_{\Be_1\otimes\Be_0}\in J$, we have $J=[K]$ by Proposition \ref{prop:sec_L}, so that
$(K^T)^\perp=(J^T)^\perp$. Note also that
$S^T=S$, and $S^*(a)=I_{\Be_1}\otimes a$ for $a\in \Be_0$. By Lemma \ref{lemma:S_chan},
\[
(K^T)^\perp=I_{\Be_1}\otimes[I_{\Be_0}]^\perp
\]
 and taking into account that  $X_\Phi\ge 0$, we get
\begin{equation}\label{eq:gchan_on_chan}
\ptr_\Be X_\Phi\in [I_{\Be_1}\otimes(\tau_{\Be_0} + [I_{\Be_0}]^\perp)] \cap (\Be_1\otimes\Be_0)^+=I_{\Be_1}\otimes \Se(\Be_0).
\end{equation}
Moreover,  $\Phi_1$ and $\Phi_2$ are equivalent if and only if
\[
X_{\Phi_1}-X_{\Phi_2}=I_{\Be_1}\otimes Y,\quad Y\in \Be\otimes\Be_0,\ \ptr_{\Be_0}Y=0
\]

\qed
\end{ex}

\begin{ex}\label{ex:chan_on_POVM} \textbf{(Channels on POVMs)} Put $\Be_1=\mathbb C^m$ in Example \ref{ex:channels}, then
$\Ce(\Be_0,\mathbb C^m)$ is the set of all POVMs on $\Be_0$, with values in $\{1,\dots,m\}$.
If $\Phi: \Be_1\otimes\Be_0\to \Be$ is a cp map, then the Choi matrix has the form $X_\Phi=\sum_{j=1}^m |j\>\<j|\otimes X_j$,
 $X_j\in (\Be\otimes\Be_0)^+$.
The condition (\ref{eq:gchan_on_chan}) becomes
\begin{equation}\label{eq:gchan_on_povm}
X_\Phi=\sum_{j=1}^m |j\>\<j|\otimes X_j,\quad \ptr_\Be X_j=\omega\ \forall j,\quad \omega\in \Se(\Be_0)
\end{equation}
and $\Phi_1$ and $\Phi_2$ are equivalent if and only if $X_{\Phi_i}=\sum_j|j\>\<j|\otimes X_{ij},\ i=1,2,$ with
\[
X_{1j}-X_{2j}= Y\ \forall j,\quad Y\in \Be\otimes \Be_0,\quad \ptr_{\Be_0} Y=0
\]

\qed
\end{ex}

\begin{ex}\label{ex:OL}  Let $\Ae=B(\Ha)$ and let $E=(E_1,\dots, E_k)$ be a POVM on $B(\Ha)$.
Then $E$ defines a channel $S_E:B(\Ha)\to \mathbb C^k$ by $a\mapsto (\Tr (E_1 a),\dots, \Tr (E_k a))$.
Let $\rho$ be a faithful state and let $S_E(\rho)=\lambda=(\lambda_1,\dots,\lambda_k)$. Let
$J=S_E^{-1}([ \lambda])$ and let
\[
K= J\cap \Se(\Ae)=
\{ \sigma\in \Se(\Ha), \, \Tr (\sigma E_i)=\lambda_i, i=1,\dots,k\}
\]
We have $S_E^T=S_{E^T}$ and $(S_E^T)^*(x)= \sum_i x_i E^T_i$ for $x\in \mathbb C^k$, 
and since $\rho\in J$ is invertible,  $(K^T)^\perp=(J^T)^\perp=S_{E^T}^*([\lambda]^\perp)$,
by Lemma \ref{lemma:S_chan}.
It follows that channels  $K\to \Se(\Be)$ are given by cp maps $\Phi:B(\Ha)\to \Be$,
such that
\[
\ptr_\Be X_\Phi=\sum_i c_i E_i^T,\qquad \sum_ic_i\lambda_i=1
\]
Note that if $E$ is a PVM, then $E^T$ is  a PVM as well and positivity of $X_\Phi$ implies that we must have $c_i\ge0$ for all $i$.
Moreover, $\Phi_1$ and $\Phi_2$ are equivalent if and only if
\[
X_{\Phi_1}-X_{\Phi_2}=\sum_j y_j\otimes E_j^T,\quad y_j\in \Be,\ \sum_j \lambda_jy_j=0
\]

More generally, let $E^i=(E^i_1,\dots,E^i_{k_i})$, $i=1,\dots,n$ be POVMs. Put $J=\cap_i J_i$, for
$J_i=S_{E^i}^{-1}([\lambda^i])$, with $\lambda^i_j=\Tr (E^i_j\rho)$,  $j=1,\dots,k_i$, $i=1,\dots,n$, and
\[
K= J \cap \Se(\Ae)=\{\sigma\in \Se(\Ae),\ \Tr(\sigma E^i_j)=\lambda^i_j,\ j=1,\dots,k_i,\ i=1,\dots,n\}
\]
Again, $\rho\in J$, so that
\[
(K^T)^\perp=(J^T)^\perp=(\cap_i J_i^T)^\perp=\vee_i (J_i^T)^\perp=\vee_i S_{E_i^T}^*([\lambda^i]^\perp)
\]
It follows that channels $K\to \Se(\Be)$ are given by cp maps $\Phi:B(\Ha)\to \Be$, satisfying
\[
\ptr_\Be X_\Phi=\sum_{i=1}^n\sum_{j=1}^{k_i} d^i_j (E^i_j)^T,\quad \sum_{i,j} d^i_j\lambda^i_j=1
\]
and $\Phi_1$, $\Phi_2$ are equivalent if and only if
\[
X_{\Phi_1}-X_{\Phi_2}=\sum_{ij} y_{ij}\otimes (E^i_j)^T,\quad y_{ij}\in \Be,\ \sum_j y_{ij}\lambda^i_j=0,\ \forall i
\]

\qed
\end{ex}


\subsection{Measurements and instruments on $K$}

Let    $\Be=\mathbb C^m\otimes B(\Ka_1)$ and let $\Phi:\Ae\to \Be$ be a
generalized channel with respect to $K$. Then there are cp maps $\Phi_j:\Ae\to B(\Ka_1)$, $j=1,\dots,m$, such that
$\Phi(a)=\sum_j |j\>\<j|\otimes \Phi_j(a)$.  Since
\[
1=\Tr(\Phi(a))=\sum_j\Tr(\Phi_j(a)),\qquad a\in K,
\]
$\sum_j \Phi_j$ is a generalized channel with respect to $K$.
In this case, we will say that $\Phi$ is a generalized instrument with respect to $K$.

In particular, let $\Be=\mathbb C^m$, then any cp map $\Phi: \Ae\to \Be$ has the form
(\ref{eq:POVM}) with  some positive elements $M_j\in \Ae$
and the Choi matrix is $X_\Phi=\sum_j|j\>\<j|\otimes M_j^T$.
Then $\Phi$ is a generalized channel with respect to $K$ if and only if
\begin{equation}\label{gPOVM}
\sum_j M_j=\ptr_\Be X^T_\Phi\in I_\Ae+ K^\perp.
\end{equation}
Any such collection of positive operators will be called a generalized POVM (with respect to $K$). If  $M$ and $N$  are
generalized POVMs, then they are equivalent if and only if
\begin{equation}\label{gPOVM_equiv}
M_j-N_j\in K^\perp,\quad \forall j
\end{equation}

Now let $K$ be any convex subset of $\Se(\Ae)$. 
A measurement on $K$ with values in a finite set $X$ is naturally defined as  an affine map
 from $K$ to the set of probability measures on $X$.
It is clear that any generalized POVM with respect to $K$ defines a measurement 
on $K$
 by
 \[
 p_j(a)=\Tr (M_j a),\quad j\in X\qquad a\in K
 \]
Conversely, any measurement on $K$  is given by a collection of functions $\lambda_i\in A(K)^+$, $i\in X$, such that $\sum_i\lambda_i=1$
 (here 1 is the function identically 1 on $K$). Each $\lambda_i$ is given by some element $M_i\in \Ae^h$, such that
 $\sum_iM_i \in I_\Ae+K^\perp$.
  By Theorem  \ref{thm:ext}, all $M_i$ can be chosen  positive, and hence form a generalized POVM,  if and only if
  $\lambda$ extends to a measurement on the section
  $\tilde K$, see  Remark \ref{rem:gch}. If $K$ is a section of $\Se(\Ae)$, then measurements on $K$ are precisely the equivalence classes of generalized POVMs. If $K$ is not a section, then Theorem \ref{thm:ext} implies that there are measurements on $K$
  that cannot be obtained by a generalized POVM.

\begin{ex}\label{ex:ppovm}\textbf{(PPOVMs)}
Let $\Be_0=\Be(\Ha_0)$, $\Be_1=B(\Ha_1)$ and $\Be=\mathbb C^k$ in Example \ref{ex:channels}. Let us denote $\Ce(\Ha_0,\Ha_1):=\Ce(\Be_0,\Be_1)$ in this case. Since
this is (a multiple of) a section of $\Se(\Ha_1\otimes \Ha_0)$, measurements on
 $\Ce(\Ha_0,\Ha_1)$ are given by generalized POVMs.
A collection $(M_1,\dots,M_m)$ of operators  $M_i\in B(\Ha_1\otimes \Ha_0)^+$
is a generalized POVM with respect to $\Ce(\Ha_0,\Ha_1)$ if and only if
\[
\sum_j M_j=I_{\Ha_1}\otimes \omega,\qquad \omega\in \Se(\Ha_0)
\]
Note that these are exactly the quantum 1-testers \cite{daria_testers}, also called 
 process POVMs, or PPOVMs, in \cite{ziman}. Moreover, two
PPOVMs $M$ and $N$ are equivalent if and only if
\[
M_j-N_j=I_{\Ha_1}\otimes y_j,\quad \Tr(y_j)=0,\ \forall j
\]
Similarly, if we put $\Be_0=B(\Ha_0)$, $\Be_1=\mathbb C^m$ and $\Be=\mathbb C^k$,
 we get that any measurement on the set $\Ce(B(\Ha_0),\mathbb C^m)$  has the form
 $(M_1,\dots,M_k)$, with
\[
M_j=\sum_{i=1}^m|i\>\<i|\otimes M_{ij},\ M_{ij}\in B(\Ha_0)^+,\quad \sum_j M_{ij}=\omega\in \Se(\Ha_0),\ \forall i
\]
and $M$ and $N$ define the same measurement if and only if
\[
M_{ij}-N_{ij}=y_j,\ \forall i,\quad \Tr (y_j)=0,\ \forall j
\]

\qed
\end{ex}

\subsection{Decomposition of generalized channels}

Let $c\in \Ae^+$. We denote $\chi_c: a\mapsto c^{1/2}ac^{1/2}$. Then $\chi_c$ is
a  completely positive map $\Ae\to \Ae$ and  $\chi_c$ defines a  channel on
$K$ if and only if $\Tr (\chi_c(a))=\Tr (a c)=1$, that is,  $\Tr((I_\Ae-c)a)=0$ for all $a\in K$. This shows that
$\chi_c$ is a generalized channel if and only if 
\[
c\in \bigcap_{\sigma\in K}\Se_\sigma(\Ae)= (I_\Ae+ K^\perp)\cap \Ae^+
\]
Such  generalized channels with respect to $K$ will be called simple.

\begin{prop}\label{prop:preproc} Let  $\Phi:\Ae\to \Be$ be a generalized channel with respect to $K$.
Then there is a pair $(\chi,\Lambda)$,
with $\chi=\chi_c$ a simple generalized channel with respect to   $K$  and $\Lambda:\Ae\to \Be$ a channel, such that
 \[
 \Phi=\Lambda\circ \chi
\]
Conversely, each such pair defines a generalized channel. If in  each pair
$(\chi,\Lambda)$ we take the restriction $\Lambda|_{\Ae_p}$ with $p=\supp(c)$,
then the correspondence is one-to-one.

\end{prop}

{\it Proof.} Let $\Phi:\Ae\to \Be$ be a generalized channel. Then
$\ptr_\Be X_\Phi \in (I_\Ae + (K^T)^\perp )\cap \Ae^+$, or, equivalently,
\[
\Phi^*(I_\Be)\in (I_\Ae+ K^\perp)\cap \Ae^+
\]
Put $c=\Phi^*(I_\Be)$
and let $p=\supp(c)$. Then since $b\le \|b\|I_\Be$ for 
$b\in \Be^+$, we have
$\Phi^*(b)\le \|b\|c\le \|b\|\|c\|p$. This implies that $p\Phi^*(b)p=\Phi^*(b)p=\Phi^*(b)$
for all  $b\in \Be^+$,
and hence for all $b\in \Be$, so that $\Phi^*$ maps $\Be$ into $\Ae_p$. It
follows that $\chi_{c^{-1}}\circ \Phi^*$ is well defined and unital map
$\Be\to \Ae_p$. Let  $\Lambda_p$ be the adjoint map,
$\Lambda_p=\Phi\circ\chi_{c^{-1}}$, then $\Lambda_p$   is a channel
$\Ae_p\to \Be$  and $\Phi=\Lambda_p\circ\chi_c$.

The channel $\Lambda_p$ can be extended to a channel $\Lambda: \Ae\to \Be$ as
\[
\Lambda(a)=\Lambda_p(a)+ \omega \Tr a(1-p),\qquad a\in \Ae
\]
where $\omega\in \Be$ is any state, and $\Phi=\Lambda\circ\chi_c$.  The converse is quite obvious.

Suppose now that there are $(\chi_i,\Lambda_i)$, $i=1,2$, such that $\Phi_1:=\Lambda_1\circ\chi_1=\Lambda_2\circ\chi_2=:\Phi_2$.
Let $\chi_i=\chi_{c_i}$.
Then since $\Phi_i^*(I_\Be)=c_i$, we have $c_1=c_2=:c$ and $\chi_1=\chi_2=:\chi$. Let $p:=\supp c$.
But then it is clear that if $\Lambda_i$ are defined on $\Ae_p$,
then we must have $\Lambda_i=\Phi\circ \chi_c^{-1}$.

\qed

We apply this result to the set of channels on $\Ce(\Ha_0,\Ha_1)$, see Example \ref{ex:channels}.

\begin{thm}\label{thm:gen_chanels_on_chans} For any channel
$\Xi : \Ce(\Ha_0,\Ha_1)\to \Se(\Be)$, there exists an ancillary Hilbert
space $\Ha_A$, a pure state $\rho\in B(\Ha_0\otimes \Ha_A)$ and a channel
$\Lambda: B(\Ha_1\otimes \Ha_A)\to \Be$, such that
\begin{equation}\label{eq:chan_on_chan}
\Xi(X_\mathcal E)=\Lambda\circ(\mathcal E\otimes id_{\Ha_A})(\rho),\qquad \mathcal E\in \Ce(\Ha_0,\Ha_1)
\end{equation}
Conversely, let $\Ha_A$ be an ancillary Hilbert space and let $\rho\in B(\Ha_0\otimes\Ha_A)$ be a state.
Let $\Lambda: B(\Ha_1 \otimes \Ha_A)\to \Be$ be a channel. Then (\ref{eq:chan_on_chan}) defines a channel
$\Ce(\Ha_0,\Ha_1)\to \Se(\Be)$.

\end{thm}

{\it Proof.} By  Example \ref{ex:channels} and Proposition \ref{prop:preproc},
\[
\Phi=\Lambda\circ\chi_{I_{\Ha_1}\otimes \omega}
\]
with  $\omega\in \Se(\Ha_0)$ and
$\Lambda: B(\Ha_1\otimes p\Ha_0)\to B(\Ka)$ a channel, $p=\supp \omega$.
Let now $\mathcal E:B(\Ha_0)\to B(\Ha_1)$ be a channel.
 Then  we have
\[
\chi_{I\otimes \omega}(X_{\mathcal E})=(I_{\Ha_1}\otimes\omega^{1/2})(\mathcal E\otimes id_{\Ha_0})(\Psi_{\Ha_0})(I_{\Ha_1}\otimes\omega^{1/2})=
(\mathcal E\otimes id_{p\Ha_0})(\rho)
\]
where $\rho= (I_{\Ha_0}\otimes\omega^{1/2})\Psi_{\Ha_0}(I_{\Ha_0}\otimes\omega^{1/2})$ is a pure state in $B(\Ha_0\otimes p\Ha_0)$.
Then (\ref{eq:chan_on_chan}) holds, with $\Ha_A=p\Ha_0$.

To prove the converse,
let $\mathcal R: B(\Ha_0)\to B(\Ha_A)$ be the cp map with Choi matrix $\rho$, then
$\rho=(id_{\Ha_0}\otimes \mathcal R)(\Psi_{\Ha_0})$.
We have
\[
(\mathcal E\otimes id_{\Ha_A})(\rho)=(\mathcal E\otimes id_{\Ha_A})(id_{\Ha_0}\otimes \mathcal R)(\Psi_{\Ha_0})=
(id_{\Ha_1}\otimes \mathcal R)(\mathcal E\otimes id_{\Ha_0})(\Psi_{\Ha_0})
\]
Put $\Phi=\Lambda\circ(id_{\Ha_1}\otimes \mathcal R)$, then $\Phi$ is a cp map $B(\Ha_1\otimes \Ha_0)\to \Be$ and
\[
\Phi^*(I_\Be)=(id_{\Ha_1}\otimes\mathcal R^*)(I_{\Ha_1\otimes \Ha_A})=I_{\Ha_1}\otimes \omega
\]
where $\omega=\mathcal R^*(I_{\Ha_A})=\ptr_{\Ha_A} \rho^T$ is a state in $B(\Ha_0)$.

\qed

Note that the analog to the above Theorem for PPOVMs was proved in \cite{ziman}.

\section{Generalized supermaps}\label{sec:super}

Quantum supermaps were defined in \cite{daria_super} as completely positive map transforming  a quantum 
operation to another quantum operation. More generally, supermaps on supermaps,
 or quantum combs, were introduced in \cite{daria_circ}. In this section, we define generalized supermaps as 
channels on generalized channels and show the relation to quantum combs.

Let   $J\subseteq \Ae$ be a self-adjoint subspace. Denote by  $\tilde J$ the
vector
subspace generated by $I_\Ae+ (J^T)^\perp$. Then it is easy to see that
$\tilde J$ is self-adjoint and
\[
\tilde J=[I_\Ae]\vee (J^T)^\perp
\]

\begin{lemma}\label{lemma:tilde}
\begin{enumerate}
\item[(i)] If $\rho\in J$ is any state, then
\[
(I_\Ae+(J^T)^\perp)\cap \Ae^+=\tilde J\cap \Se_{\rho^T}(\Ae)
\]
\item[(ii)] If $I_\Ae\in J$, then $\tilde{\tilde J}=J$.
\item[(iii)] If  $J=S^{-1}(J_0)$ for  channel $S:\Ae\to \Ae_0$ and a self-adjoint subspace $J_0\subseteq \Ae_0$, then $\tilde J=(S^T)^*(\tilde J_0)$.
\end{enumerate}
\end{lemma}

{\it Proof.} (i) An element $x\in \tilde J$ has the form
$x=c I_\Ae+x_0$, where $x_0\in (J^T)^\perp$ and  $c=\Tr \rho^Tx$
for any state $\rho\in J$.
(ii) follows from the fact that if   $I_\Ae\in J$, then
\[
\tilde{\tilde J}=[I_\Ae]\vee (\tilde J^T)^\perp=[I_\Ae]\vee ([I_\Ae]^\perp\wedge J)=J,
\]
(iii) follows from Lemma \ref{lemma:S_chan}.
\qed

Let $K$ be a section of $\Se(\Ae)$ and let $J=[K]$. We denote by
$\Ce_K(\Ae,\Be)$ or $\Ce_J(\Ae, \Be)$
the set of all generalized channels $\Ae\to \Be$ with respect to $J$.  In
particular,  if $K=\Se(\Ae)$, we get the set of all channels  $\Ce(\Ae, \Be)$.
An element  $\Phi\in \Ce_J(\Ae, \Be)$ will
 be identified with its  Choi matrix $X_\Phi\in \Be\otimes \Ae$.
In the next Proposition, we characterize the set $\Ce_J(\Ae,\Be)$.

\begin{prop}\label{prop:set_of_gchans}  Let $K$ be a section of $\Se(\Ae)$
and let $J= [K]$. Then
\[
\Ce_J(\Ae,\Be)=\ptr_\Be^{-1}(\tilde J)\cap \Se_{I_\Be\otimes \rho^T}
(\Be\otimes \Ae)
\]
where $\rho$ is any element in $K$.
In particular, if $K$ contains the tracial state $\tau_\Ae$,
then $\Ce_J(\Ae, \Be)=
 \ptr_\Be^{-1}(\tilde J)\cap t_\Ae\Se(\Be\otimes \Ae)$.

\end{prop}

{\it Proof.} An element $X\in \Be\otimes \Ae$ is the Choi matrix
of a generalized channel with respect to $J$
if and only if  $X$ is positive and
\[
\ptr_\Be X \in (I_\Ae+ (J^T)^\perp)\cap \Ae^+= \tilde J\cap\Se_{\rho^T}(\Ae),
\]
 by Lemma \ref{lemma:tilde} (i), which is equivalent with $\ptr_\Be X\in \tilde J$ and
$1=\Tr{\rho^T} \ptr_\Be X=\Tr (I_\Be\otimes {\rho^T})X$.

If $\tau_\Ae\in K$, then
$\Se_{I_\Be\otimes \tau_\Ae^T}(\Be\otimes\Ae)=t_\Ae\Se(\Be\otimes\Ae)$.

\qed

This implies that if $K$ contains the tracial state,
then the set of generalized channels forms
a constant multiple of a section of the state space $\Se(\Be\otimes\Ae)$.
Then any cp map that maps $C_J(\Ae,\Be)$ to another state space is a constant
multiple of a generalized channel.
Since the set $\ptr_\Be^{-1}(\tilde J)$ always contains the unit,
we can repeat the process infinitely. The  generalized channels
obtained in this way will be called generalized supermaps.

Let $\Be_0,\Be_1,\Be_2,\dots $ be   finite dimensional $C^*$ algebras
and let $K$ be a section of the state space $\Se(\Be_0)$, such that $\tau_{\Be_0}\in K$.
Let $J=[K]$. We  denote by
$\Ce_J(\Be_0,\Be_1,\dots,\Be_n)$ the set of all cp maps that map
$\Ce_J(\Be_0,\Be_1,\dots,\Be_{n-1})$ into $\Se(\Be_n)$.
We further introduce the following notations. Let
$\Ae_n:=\Be_n\otimes\Be_{n-1}\otimes\dots\otimes\Be_0$, $n=0,2,\dots$. Let $S_n :\Ae_n\to \Ae_{n-1}$ denote the partial trace
$\Tr^{\Ae_n}_{\Be_n}$, $n=1,2,\dots$.

\begin{thm}\label{thm:gchansongchans} We have for $n=1,2,\dots,$
\[
\Ce_J(\Be_0,\dots,\Be_n)=J_n\cap c_n \Se(\Ae_n)
\]
where
\begin{eqnarray*}
J_{2k-1} &=& J_{2k-1}(J,\Be_1,\dots,\Be_{2k-1}):= S_{2k-1}^{-1}(S_{2k-2}^*(S_{2k-3}^{-1}(\dots S_1^{-1}(\tilde J)\dots)))\\
J_{2k} &=& J_{2k}(J,\Be_1,\dots,\Be_{2k}):=S_{2k}^{-1}(S_{2k-1}^*(S_{2k-2}^{-1}(\dots S_1^*(J)\dots)))
\end{eqnarray*}
and $c_n=c_n(J,\Be_1,\dots,\Be_{2k-1}):=\Pi_{l=0}^{\lfloor \frac{n-1}2\rfloor}
t_{\Be_{n-1-2l}}$.

\end{thm}

{\it Proof.} We will prove the statement by induction on $n$, together with the
fact that $J_n=S_n^{-1}(\tilde J_{n-1})$ for $n=1,2,\dots$,  where we put $J_0:=J$.

For   $n=1$, the statement is  proved
 in Proposition \ref{prop:set_of_gchans} and $J_1=S_1^{-1}(\tilde J)$ by
definition.
Suppose now that this  holds for some
$n$. Note that since $\tilde J_{n-1}$ contains the unit $I_{\Ae_{n-1}}$,
$J_n=S_n^{-1}(\tilde J_{n-1})$ contains the unit as well. Then
\begin{equation}\label{eq:Ce_K}
\Ce_J(\Be_0,\dots,\Be_{n+1}) =\frac1{c_n}\Ce_{J_n}(\Ae_n,\Be_{n+1})
\end{equation}
and by  Proposition \ref{prop:set_of_gchans},
\[
\Ce_{J_n}(\Ae_n,\Be_{n+1}) = S_{n+1}^{-1}(\tilde J_n)\cap t_{\Ae_n} \Se(\Ae_{n+1})
\]
Since $S_n^T=S_n$, we have by Lemma \ref{lemma:tilde} (ii) and (iii)
that
\begin{equation}\label{eq:tildeJ}
\tilde J_n=S_n^*(\tilde{\tilde {J}}_{n-1})=
S_n^*(J_{n-1})
\end{equation}
so that $S_{n+1}^{-1}(\tilde J_n)=J_{n+1}$. Finally, the proof follows from
\[
\frac{t_{\Ae_n}}{c_n}=\frac{\Pi_{l=0}^n t_{\Be_l}}{\Pi_{l=0}^{\lfloor \frac{n-1}2\rfloor}
t_{\Be_{n-1-2l}}}=\Pi_{l=0}^{\lfloor \frac{n}2\rfloor}
t_{\Be_{n-2l}} = c_{n+1}.
\]

\qed

The above theorem can be written in the following form:

\begin{thm}\label{thm:gchans_daria} Let  $k:=\lfloor \frac n2\rfloor$.
Then $X\in \Ce_J(\Be_0,\dots,\Be_n)$ if and only if there are positive
elements $Y^{(m)}\in \Ae_{n-2m}$ for
$m=0,\dots,k$, such that
\begin{equation}\label{eq:daria}
\ptr_{\Be_{n-2m}} Y^{(m)}=I_{\Be_{n-2m-1}}\otimes Y^{(m+1)}, m=0,\dots,k-1
\end{equation}
 $Y^{(0)}:=X$,  $Y^{(k)} \in \Ce_J(\Be_0,\Be_1)$ if $n=2k+1$
and $Y^{(k)}\in K$ if $n=2k$.

\end{thm}

\begin{ex}\label{ex:chan_on_gPOVM}
(\textbf{Channels on generalized POVMs}) Let $X\in
\Ce_J(\Ae,\mathbb C^m,\Be)$, then $X$ defines a channel on the set
$\Ce_J(\Ae,\mathbb C^m)$ of generalized POVMs. Since $X\in \Be\otimes \mathbb C^m\otimes \Ae$, we must have
$X=\sum_{j=1}^m |j\>\<j|\otimes X_j$, $X_j\in \Be\otimes\Ae$. By Theorem
\ref{thm:gchans_daria}, $\ptr_{\Be} X =I_{\mathbb C^m}\otimes X_0$ for some
$X_0\in K$. It follows that if $X$ is positive,
\begin{equation}\label{eq:chan_on_POVM}
X\in \Ce_J(\Ae,\mathbb C^m,\Be)\iff   X=\sum_{j=1}^m |j\>\<j|\otimes X_j,\ \ptr_\Be X_j=X_0\in K,\ \forall j
\end{equation}

\qed
\end{ex}

Note that Example \ref{ex:chan_on_POVM} is a special case of the above example. Another special case is the following:

\begin{ex}\label{ex:chan_on_PPOVM} (\textbf{Channels and measurements on PPOVMs})
 Let $\Ha_0$, $\Ha_1$ be finite dimensional Hilbert spaces. Then $\Ce(B(\Ha_0),B(\Ha_1),\mathbb C^m)$ is the set of all
measurements on $\Ce(\Ha_0,\Ha_1)$ with values in $\{1,\dots,m\}$, that is,
 the set of all PPOVMs. By (\ref{eq:Ce_K}),
\[
\Ce(B(\Ha_0),\Be(\Ha_1),\mathbb C^m)=\frac 1{\dim\Ha_0}\Ce_{J_1}(B(\Ha_1\otimes\Ha_0),\mathbb C^m)
\]
so that
\[
\Ce(B(\Ha_0),B(\Ha_1),\mathbb C^m,\Be)=(\dim \Ha_0) \Ce_{J_1}(B(\Ha_1\otimes\Ha_0),\mathbb C^m,\Be)
\]
here $J_1=\ptr_{\Ha_1}^{-1}([I_{\Ha_0}])$. By (\ref{eq:chan_on_POVM}),
$X\in \Ce(B(\Ha_0),B(\Ha_1),\mathbb C^m,\Be)$ if and only if
\[
X=\sum_{j=1}^m |j\>\<j| \otimes X_j,\quad  \ptr_\Be X_j =X_0\in \Ce(B(\Ha_0),B(\Ha_1)),\ \forall j
\]
Note that by Theorem \ref{thm:combs} below, this also describes all cp maps sending
 POVMs with values in $\{1,\dots,m\}$ to channels $B(\Ha_0)\to \Be$.

In particular, by putting $\Be=\mathbb C^k$, we get that measurements on PPOVMs
are given by collections of instruments $\Lambda_j:B(\Ha_0)\to B(\Ha_1)$ with values in  $\{1,\dots,k\}$,
such that their components $\Lambda_{1j},\dots, \Lambda_{kj}$
sum to the same channel, for all $j\in\{1,\dots,m\}$.

\end{ex}

Let now $K=\Se(\Be_0)$. Then $J=\Be_0$ and $\tilde J= [I_{\Be_0}]$, so that
Proposition \ref{prop:set_of_gchans} gives the usual characterization
of the set $\Ce(\Be_0,\Be_1)$ of all Choi matrices of
channels  $\Be_0\to\Be_1$. For $n>1$,  we have the characterization in Theorem \ref{thm:gchans_daria},
 with $Y^{(k)}\in \Se(\Be_0)$ if $n=2k$ and
$\ptr_{\Be_1}Y^{(k)}=I_{\Be_0}$ for $n=2k+1$. Suppose that all $\Be_j$, $j=0,1,\dots,$ are matrix algebras, $\Be_j=B(\Ha_j)$. Then, comparing Theorem
\ref{thm:gchans_daria} with the results in \cite{daria}, we see that for
$n=2k-1$, the set
$\Ce(B(\Ha_0),\dots,B(\Ha_n))$ is precisely the set of $k$ -combs on
$(\Ha_0,\dots,\Ha_{2k-1})$. We give the definition below and also give an
alternative proof of the characterization of quantum  combs. Note that a similar characterization was obtained in \cite{guwat} for Choi matrices of strategies and 
co-strategies of quantum games.

\subsection{Quantum combs}

Quantum $N$- combs were defined in \cite{daria} as a tool for description of 
quantum networks.  
A quantum 1-comb on $(\Ha_0,\Ha_1)$  is the Choi matrix of
a channel $B(\Ha_0)\to B(\Ha_1)$. A quantum
$N$-comb on $(\Ha_0,\Ha_1,\dots,\Ha_{2N-1})$ is the Choi matrix of a cp map,
transforming $(N-1)$-combs on $(\Ha_1,\dots,\Ha_{2N-2})$ to 1-combs on
$(\Ha_0,\Ha_{2N-1})$. We use the definition of $N$-combs with 
the matrix algebras $B(\Ha_j)$ replaced by
 finite dimensional $C^*$-algebras $\Be_j$, $j=0,\dots,2N-1$. This corresponds 
to   conditional combs introduced in \cite{daria_bit}, 
 which describe
quantum networks with classical inputs and outputs.
We   show below that the $N$-combs are precisely the generalized supermaps
$\Ce(\Be_0,\dots, \Be_{2N-1})$.

Let $\Ae, \Be,\Ce$ be finite dimensional $C^*$-algebras and let $K$ be a section
of $\Se(\Ae)$, let $J=[K]$. We will describe
 the set of all cp maps $\Ae\to \Ce\otimes \Be$ that transform  $K$ into
the set of all channels
$\Be\to \Ce$, this will be denoted by $\comb_J(\Ae,\Be,\Ce)$. It will be
convenient
to consider this set as a subset in $\Ce\otimes\Ae\otimes\Be$.

It is quite
 clear that if $X\in (\Ce\otimes\Ae\otimes\Be)^+$, then
$X\in \comb_J(\Ae,\Be,\Ce)$ if and only if
$X*\rho \in C_J(\Ae,\Ce)$ for all $\rho\in \Se(\Be)$, this follows from
(\ref{eq:link_value}) and from
\[
 (X*\rho)*a= X*(a\otimes \rho)= (X*a)*\rho,
 \]
 for all $\rho\in \Se(\Be)$ and $a\in K$.

\begin{prop}\label{prop:combs} Suppose  $\tau_\Ae\in K$.  Then
\[
\comb_J(\Ae,\Be,\Ce)=\Ce_{J\otimes\Be }(\Ae\otimes \Be,\Ce)
\]

\end{prop}

{\it Proof.} Let $X$ be a positive element in $\Ce\otimes\Ae\otimes\Be$.
As we already argued above, $X\in \comb_J(\Ae,\Be,\Ce)$ if and
 only in $X*\rho\in C_J(\Ae,\Ce)$ for all $\rho\in \Se(\Be)$, in other words,
\begin{equation}\label{eq:comb_1}
\ptr_{\Ce}(X*\rho)=(\ptr_{\Ce}X)*\rho\in \tilde J,\qquad \rho\in\Se(\Be)
\end{equation}
and, simultaneously,
\begin{equation}\label{eq:comb_2}
\Tr(X*\rho)=\Tr (\rho^T[\ptr_{\Ce\otimes\Ae}X])=t_\Ae, \qquad \rho\in \Se(\Be),
\end{equation}
which means that $\ptr_{\Ce\otimes\Ae}X=t_\Ae I_\Be$.
Moreover, we can write (\ref{eq:comb_1}) as
\[
0=\Tr [((\ptr_{\Ce}X)*\rho)a]=\Tr[(\ptr_{\Ce}X)(a\otimes \rho^T)]
\]
for all $\rho\in \Be$ and $a\in \tilde J^\perp$, which is the same as
$\ptr_{\Ce}X \in (\tilde J^\perp\otimes \Be)^\perp= \tilde J\otimes \Be$. Putting this together, we get
$X\in \comb_J(\Ae,\Be,\Ce)$ if and only if
\[
\ptr_\Ce X\in [\tilde J\otimes\Be]\wedge S_\Ae^{-1}([I_\Be]),\qquad \Tr X=t_{\Ae\otimes\Be},
\]
where $S_\Ae:=\ptr_\Ae^{\Ae\otimes \Be}$.

Let $Y\in \tilde J\otimes \Be$, then $Y=\sum_i (t_i I_\Ae+x_i)\otimes b_i$, with $b_i\in \Be$ and $x_i\in (J^T)^\perp$.
Since $\tau_\Ae\in K$, we have $\ptr_\Ae Y= t_\Ae\sum_it_ib_i$, so that $\ptr_\Ae Y\in [I_\Be]$ if and only if
$Y=cI_{\Ae\otimes\Be}+
\sum_i x_i \otimes b_i$ for some $c\in \mathbb C$, this implies that
\[
Y\in [I_{\Ae\otimes\Be}]\vee ((J^T)^\perp\otimes \Be)=
(J\otimes \Be)\tilde{\ }.
\]
Conversely, let $Y\in(J\otimes \Be)\tilde{\ }$ and let $\{b_k\}_k$ be a basis in $\Be$, such that $b_1=I_\Be$. Then there are
 $x_k\in (J^T)^\perp$, such that
 $Y=cI_{\Ae\otimes\Be}+\sum_k x_k\otimes b_k= \sum_k (t_k I_\Ae+x_k)\otimes b_k$, with $t_1=c$ and $t_k=0$ for $k\ne 1$. Hence
  $Y\in  \tilde J\otimes \Be$, and, clearly, $\ptr_\Ae Y\in[I_\Be]$.
This proves that $[\tilde J\otimes\Be]\wedge S_\Ae^{-1}([I_\Be])=(J\otimes\Be)\tilde{\ }$, so that by Proposition \ref{prop:set_of_gchans},
\[
\comb_J(\Ae,\Be,\Ce)=\ptr_\Ce^{-1}((J\otimes\Be)\tilde{\ })\cap t_{\Ae\otimes \Be}\Se(\Ce\otimes\Ae\otimes\Be)=\Ce_{J\otimes\Be}(\Ae\otimes\Be,\Ce)
\]

\qed

Let us now denote by $\comb(\Be_0,\dots,\Be_{2N-1})$ the set of $N$-combs.

\begin{thm}\label{thm:combs}$\comb(\Be_0,\dots,\Be_{2N-1})=\Ce(\Be_0,\dots,
\Be_{2N-1})$.

\end{thm}

{\it Proof.} For $N=1$, the statement is trivial.
 Suppose that it is true for some $N$.
 Let
$\hat \Ae_{2N-1}:=\Be_{2N}\otimes\dots\otimes\Be_1$ and let $\hat J_{2N-1}:=J_{2N-1}(\Be_1,\dots,\Be_{2N})$
 and $\hat c_{2N-1}=c_{2N-1}(\Be_1,\dots,\Be_{2N})$, with the notations from  Theorem \ref{thm:gchansongchans}.
Then
\begin{equation}\label{eq:combs_subspace}
\comb(\Be_1,\dots,\Be_{2N})=\Ce(\Be_1,\dots,\Be_{2N})=\hat J_{2N-1}\cap \hat c_{2N-1}\Se(\hat \Ae_{2N-1})
\end{equation}
Next, let $\Ae_{2N}=\hat \Ae_{2N-1}\otimes \Be_0$,
$J_{2N}=J_{2N}(\Be_0,\dots, \Be_{2N})$ and $c_{2N}=c_{2N}(\Be_0,\dots,\Be_{2N})$. Then it is not
difficult to see that $J_{2N}=\hat J_{2N-1}\otimes \Be_0$ and $c_{2N}=\hat c_{2N-1}$.
By (\ref{eq:combs_subspace}) and Proposition \ref{prop:combs}
\begin{eqnarray*}
\comb(\Be_0,\dots,\Be_{2N+1})&=&\frac 1{\hat c_{2N-1}}\comb_{\hat J_{2N-1}}(\hat \Ae_{2N-1},\Be_0,\Be_{2N+1})\\
&=&\frac 1{c_{2N}}\Ce_{J_{2N}}(\Ae_{2N},\Be_{2N+1})\\
&=&\Ce(\Be_0,\dots,\Be_{2N+1}),
\end{eqnarray*}
the last equality follows from (\ref{eq:Ce_K}).

\qed

In accordance with this result, the elements in $\Ce_J(\Be_0,\dots,\Be_{2N-1})$
 will be called generalized $N$-combs.

Note that an element   $X\in \Ce(\Be_0,\dots,\Be_{2N-1})$ is the Choi matrix of
a generalized supermap $\Be_{2N-2}\otimes\dots\otimes \Be_0\to \Be_{2N-1}$,
whereas the same operator as an element in $\comb(\Be_0,\dots,\Be_{2N-1})$
is viewed as the Choi matrix of a cp map $\Be_{2N-2}\otimes\dots\otimes\Be_1\to
\Be_{2N-1}\otimes\Be_0$. Note also that the set $\Ce(\Be_0,\dots,\Be_{2N-1},\mathbb C^k)$ is precisely the set of $N$-testers with $k$ values \cite{daria_testers},
 so that quantum testers are a special class of generalized POVMs.

\subsection{Decomposition of generalized supermaps}

Let $k=\lfloor \frac n2\rfloor$.  Let
 us write the  algebra $\Ae_n$ as
\begin{equation}\label{eq:rewrite_An}
\Ae_n=\Be'_{2k}\otimes\Be'_{2k-1}\otimes\dots\otimes\Be'_0,
\end{equation}
where  $\Be'_j=\Be_j$ for $j=0,\dots,n$ if $n=2k$, and $\Be'_j=\Be_{j+1}$ for
$j=1,\dots,2k$ and
$\Be'_0=\Be_1\otimes\Be_0$ if $n=2k+1$. Further,
let us suppose that $\Be'_j=\oplus_{l=1}^{n_j}B(\Ha_{B^j_l})$,
with minimal central projections
$\{q^j_{k_j}\}$, $j=0,1,\dots,2k$. Let us denote
\[
\Ie_k:=\{I=(I_{2k},\dots,I_0)\in \mathbb N^{2k+1},\ I_j\in\{1,\dots,n_j\},\
j=0,\dots,2k\}
\]
be the set of multiindices. For $I\in \Ie_k$ and $l\le k$, we  denote
$I^l=(I_{2l},\dots, I_0)\in \Ie_l$. Let
 $q(I):=\otimes_{l=0}^{2k} q_{I_{2k-l}}^{2k-l}$ and
$\Ha_{B(I)}:=\Ha_{B^{2k}_{I_{2k}}\dots
 B^0_{I_{0}}}$, then
\[
\Ae_n=\bigoplus_{I\in \Ie_k} \Ha_{B(I)}
\]
 and $q(I)$ are the minimal central projections in $\Ae_n$.

\begin{thm}\label{thm:daria} Let $X\in \Ce_J(\Be_0,\dots,\Be_{n})$. Let $k=\lfloor \frac n2\rfloor$.
Then there are:
\begin{enumerate}
\item
an ancillary Hilbert space $\Ha_D=\Ha_{D_0}=\Ha_{D_1}=\dots =\Ha_{D_{k}}$
\item  elements $X_m(I^{m-1})\in \Ce( \Be'_{2m-1}\otimes
B(\Ha_{D_{m-1}}), B(\Ha_{D_m})\otimes\Be'_{2m})$
for $m=1,\dots, k$ and for
every multiindex $I\in \Ie_k$,
\item a state $X_0\in B(\Ha_{D_0})\otimes J$ if $n=2k$, or a generalized channel
$X_0\in \Ce_J(\Be_0,B(\Ha_{D_0})\otimes\Be_1)$ if $n=2k+1$
 \end{enumerate}
such that, for all $I\in \Ie_k$,
 \begin{equation}\label{eq:links}
q(I)X=I_{D_k}*X_k(I^{k})*\dots*X_1(I^1)*X_0(I_0)
 \end{equation}
where
\begin{equation}\label{eq:dec_chan}
X_m(I^{m}):=(I_{D_m}\otimes q^{2m}_{I_{2m}}\otimes q^{2m-1}_{I_{2m-1}}\otimes
I_{D_{m-1}})X_m{(I^{m-1})}, m=1,\dots,k
\end{equation}
and $X_0(I_0)=(I_{\Ha_{D_0}}\otimes q^0_{I_0})X_0$.

\end{thm}

{\it Proof.} We proceed by induction on $k$. If $k=0$, then we must have $n=1$ and the statement is trivial.
Suppose now that the Theorem holds for some $k$.

Let  $n$ be such that $\lfloor \frac n2\rfloor=k+1$. Then
$\Ae_n=\Be'_{2k+2}\otimes\Be'_{2k+1}\otimes \Ae_{n-2}$ and
by Theorem \ref{thm:gchans_daria},
$X\in \Ce_J(\Be_0,\dots,\Be_n)$ if and only if $X$ is positive and there
is some $Y^{(1)}\in \Ce_J(\Be_0,\dots,\Be_{n-2})$ such that
\[
\ptr_{\Be'_{2k+2}} X= I_{\Be'_{2k+1}}\otimes Y^{(1)}
\]
Now by Theorem \ref{thm:apndx} from the Appendix, the last equation holds if and
only if  there is an
ancillary Hilbert space $\Ha_{D}=\Ha_{D_k}=\Ha_{D_{k+1}}$ and
\[
X_1(I_{2k+1},\Pi_j I_j^{k})\in \Ce(B(\Ha_{B^{2k+1}_{I_{2k+1}}D_k}),\Be'_{2k+2}),\quad X_0(\Pi_jI^k_j)\in B(\Ha_{D_kB(I^{k})})
\]
 with
\begin{equation}\label{eq:dec_x0}
\ptr_{D_k}X_0(\Pi_jI^{k}_j)=q(I^{k})Y^{(1)}
\end{equation}
such that
\[
(I_{\Be'_{2k+2}}\otimes q^{2k+1}_{I_{2k+1}}\otimes q(I^k))X= X_1(I_{2k+1},\Pi_j I_j^{k})*X_0(\Pi_jI^k_j)
\]
for any multiindex $I\in \Ie_{k+1}$. Put
\[
X_{k+1}(I^{k}):= \omega_{D_{k+1}}\otimes \left(\bigoplus_{i=1}^{n_{2k+1}}X_1(i,\Pi_j I_j^{k})\right)
\]
with an arbitrary state  $\omega_{D_{k+1}}\in B(\Ha_{D_{k+1}})$. Then
 $X_{k+1}(I^{k})\in \Ce( \Be'_{2k+1}\otimes
B(\Ha_{D_{k}}), B(\Ha_{D_{k+1}})\otimes\Be'_{2k+2})$ and
\[
q(I)X=I_{D_{k+1}}*X_{k+1}(I^{k+1})*X_0(\Pi_jI^k_j),
\]
where $X_{k+1}(I^{k+1})$ is given by (\ref{eq:dec_chan}).
Let now $X'_k:=\bigoplus_{J\in \Ie_k} X_0(\Pi_jJ_j)\in B(\Ha_{D_k})\otimes \Ae_{n-2}$. Then by
 (\ref{eq:dec_x0}) and $Y^{(1)}\in\Ce_J(\Be_0,\dots,\Be_{n-2})$, we get
 \[
\ptr_{B(\Ha_{D_k})\otimes \Be_{n-2}}X'_k=\ptr_{\Be_{n-2}}Y^{(1)}=I_{\Be_{n-3}}\otimes Y^{(2)}, \quad
Y^{(2)}\in \Ce_J(\Be_0,\dots,\Be_{n-4})
 \]
 which is
equivalent with $X_k'\in \Ce_J(\Be_0,\dots,B(\Ha_{D_k})\otimes\Be_{n-2})$.
 Since $\lfloor \frac{n-2}2\rfloor=k$,
 we may apply the induction hypothesis to $X_k'$.
Hence there is some ancilla $\Ha_{E}=\Ha_{E_0}=\dots=\Ha_{E_k}$, elements
$X_m(J^{m-1})\in \Ce( \Be'_{2m-1}\otimes
B(\Ha_{E_{m-1}}), B(\Ha_{E_m})\otimes\Be'_{2m})$
for $m=1,\dots, k-1$, an element  $X''_k(I^{k-1})\in \Ce( \Be'_{2k-1}\otimes
B(\Ha_{E_{k-1}}),B(\Ha_{E_kD_k})\otimes \Be'_{2k})$ and $X_0\in \Be_0$ satisfying 3.,
such that for every $J\in\Ie_k$,
\[
X_0(\Pi_jJ_j)=q(J)X'_k=I_{E_k}*X^{''}_k(J)*\dots *X_0(I_0)
\]
Note also that we may suppose $\Ha_E=\Ha_D$, exactly as in the proof of Theorem \ref{thm:apndx}.
By putting $X_k(J)=I_{E_k}*X^{''}_k(J)$, we obtain the result.

\qed

Theorem \ref{thm:daria}, together with Proposition \ref{prop:preproc}, give the following Corollary:

\begin{coro} For $k\ge 1$ and for any generalized $k$-comb
$X\in \Ce_J(\Be_0,\dots,\Be_{2k-1})$, there exist a pair
$(\chi,\Lambda)$, where $\chi: \Be_0\to \Be_0$ is a simple
generalized channel with respect to $J$ and
$X_\Lambda\in \comb(\Be_0,\dots,\Be_{2k-1})$, such that
\[
\Phi_X=\Lambda\circ(id_{\Be_{2k-1}\otimes\dots\otimes \Be_1}\otimes \chi).
\]
 Conversely, each such pair defines an element in    $\Ce_J(\Be_0,\dots,\Be_{2k+1})$.
In particular, $\Ce_J(\Be_0,\dots,\Be_{2k+1})$ is the set of cp maps
 sending $\Ce(\Be_1,\dots,\Be_{2k})$ to the set of generalized channels
 $\Ce_J(\Be_0,\Be_{2k+1})$.

\end{coro}

We will now describe how an element $Y\in \Ce_J(\Be_0,\dots,\Be_{n+1})$ acts
 on $X\in \Ce_J(\Be_0,\dots,\Be_{n})$. Let
$\Phi_Y:\Ce_J(\Be_0,\dots,\Be_{n})\to \Be_{n+1}$ be the cp map with Choi matrix $Y$.
By (\ref{eq:link_value}),
\begin{eqnarray*}
\Phi_Y(X)&=&Y*X=\ptr_{\Ae_n}[(I_{\Be_{n+1}}\otimes X^T)Y]\\
&=&\ptr_{\Ae_n}[(I_{\Be_{n+1}}\otimes \bigoplus_I q(I)X^T)\bigoplus_{i,J}(q^{n+1}_i\otimes q(J))Y]\\
&=& \ptr_{\Ae_n}[\bigoplus_{i,I}[(I_{\Be_{n+1}}\otimes q(I)X^T)(q^{n+1}_i\otimes
q(I))Y]\\
&=& \bigoplus_i \sum_I \ptr_{B(I)}[(I_{\Be_{n+1}}\otimes (q(I)X)^T)(q^{n+1}_i
\otimes q(I))Y]\\
&=&  \bigoplus_i \sum_I ((q^{n+1}_i
\otimes q(I))Y)* (q(I)X)
\end{eqnarray*}

Let now $n=2k$, so that $\lfloor \frac n2\rfloor=\lfloor \frac {n+1}2\rfloor =k$.
Then
\begin{eqnarray*}
q(I)X&=&I_{D_k}*X_k(I^{k})*\dots*X_1(I^1)*X_0(I_0)\\
(q^{n+1}_i\otimes q(I))Y &=&I_{E_k}*Y_k(\bar I^{k})*\dots*Y_1(\bar I^1)*Y_0(\bar I_0)
\end{eqnarray*}
$\bar I$ is he multiindex in $\Ie_k$, such that
$\bar I_{2k}=i$, $\bar I_j=I_{j+1}$, $j=1,\dots,2k-1$ and $\bar I_0=I_0I_1$.
Then
\[
 ((q^{n+1}_i\otimes q(I))Y)* (q(I)X)=I_{D_kE_k}*Y_k(\bar I^{k})*X_k(I^{k})*\dots
*Y_0(\bar I_0)*X_0(I_0),
\]
this follows from Proposition \ref{prop:link}, 1. and 2.
More explicitly,  we first apply the components of the channel $Y_0(\bar I_0)$
to the part of $X_0(I_0)$ in $\Be_0$, then on the part of the result in $\Be_1$, we apply the components of the channel $X_1(I_1)$, etc., both ancillas are traced out at the end.

Similarly, if $n=2k+1$, then $\lfloor \frac {n+1}2\rfloor =k+1$ and
\[
(q^{n+1}_i\otimes q(I))Y =I_{E_{k+1}}*Y_{k+1}(\hat I^{k+1})*\dots*Y_1(\hat I^1)*
Y_0(\hat I_0)
\]
where $\hat I\in \Ie_{k+1}$ is such that $\hat I_{2k+2}=i$, $\hat I_j=I_{j-1}$ for
$j=2,\dots 2k+1$ and $I_0=\hat I_1\hat I_0$. Then
\[
 ((q^{n+1}_i\otimes q(I))Y)* (q(I)X)=I_{D_kE_{k+1}}*Y_{k+1}(\hat I^{k+1})*
X_k(I^{k})*\dots
*Y_1(\hat I_1)*X_0(I_0)*Y_0(\hat I_0)
\]
Note that here  $X_0(I_0)$ is a channel, which we apply to
$Y_0(\hat I_0)$, etc.

\begin{ex}\textbf{(PPOVMs)} Let $Y\in \Ce(B(\Ha_0),B(\Ha_1),\mathbb C^m)$.
By  Theorem \ref{thm:daria}, there is some ancilla $\Ha_D$, a POVM
$M(=I_{D_1}*Y_1)\in \Ce(B(\Ha_1\otimes\Ha_{D}), \mathbb C^m)$ and
 a state $\rho(=Y_0)\in B(\Ha_{D}\otimes \Ha_0)$, such that
$Y=M*\rho$. For any  $X\in \Ce(\Ha_0,\Ha_1)$, we have
\[
Y*X=M*X*\rho=\bigoplus_{i=1}^m\Tr M_i (id_{D}\otimes \Phi_X)(\rho)
\]
where $M=(M_1,\dots, M_m)$,
compare this to Theorem \ref{thm:gen_chanels_on_chans}.
We will write such  decomposition as $Y=(\Ha_D,(M_1,\dots,M_m),\rho)$.

Next, let  $Z\in \Ce(B(\Ha_0),B(\Ha_1),\mathbb C^m,B(\Ha_3)\otimes \mathbb C^l)$, which is the set of all instruments from PPOVMs to $B(\Ha_3)$, with values in $\{1,\dots,l\}$. Then there is an ancilla $\Ha_E$ a channel
$\xi \in \Ce(B(\Ha_0),B(\Ha_{E}\otimes\Ha_1))$ and an instrument
$\Lambda \in \Ce(\mathbb C^m\otimes B(\Ha_E), B(\Ha_3)\otimes \mathbb C^l)$, such that
\[
Z=\Lambda*\xi
\]
Here $\Lambda=\oplus_{j=1}^m\Lambda_j$, where each $\Lambda_j:B(\Ha_E)\to B(\Ha_3)\otimes \mathbb C^l$ is an instrument, with components $(\Lambda_{1j},\dots,\Lambda_{lj})$.
We write $Z=(\Ha_E,(\Lambda_1,\dots,\Lambda_m),\xi)$.
Let now $Y=(\Ha_D,(M_1,\dots,M_m),\rho)$ be a PPOVM. We have
\begin{eqnarray*}
Z*Y&=&\bigoplus_i\sum_j\Lambda_{ij}(\ptr_{\Ha_D\otimes \Ha_1} (I_E\otimes M_j)
(id_D\otimes \xi)(\rho))\\
&=& \bigoplus_i\sum_j\ptr_{\Ha_D\otimes \Ha_1}(M_j\otimes I_{\Ha_3})
(id_D\otimes[(\Lambda_{ij}\otimes id_{\Ha_1})\circ\xi])(\rho)\\
&=& \bigoplus_i\sum_j\ptr_{\Ha_D\otimes \Ha_1}(M_j\otimes I_{\Ha_3})
(id_D\otimes \hat \Lambda_{ij})(\rho)
 \end{eqnarray*}
where $\hat \Lambda_j:= (\Lambda_{j}\otimes id_{\Ha_1})\circ\xi$ is an instrument $B(\Ha_0)\to B(\Ha_3\otimes \Ha_1)$, with values in $\{1,\dots,l\}$, such that
$\sum_i \ptr_{\Ha_3}\circ \hat \Lambda_{ij}  =\ptr_E\circ\xi$ for all $j$, compare this with
 Example \ref{ex:chan_on_PPOVM}.

\end{ex}

\begin{ex}\textbf{(Supermaps on instruments)}
We next describe the set \linebreak $\comb(B(\Ha_0),\Be(\Ha_1),\mathbb C^m\otimes B(\Ha_2),B(\Ha_3))$, that is, the set of cp maps from instruments $B(\Ha_1)\to B(\Ha_2)$ to
channels $B(\Ha_0)\to B(\Ha_3)$. By Theorems \ref{thm:combs} and \ref{thm:daria},
 for any such map, there is an ancillary Hilbert space $\Ha_D$, channels
$\xi_j: B(\Ha_D\otimes\Ha_2)\to B(\Ha_3)$, $j=1,\dots,m$ and a channel
$\xi: B(\Ha_0)\to B(\Ha_D\otimes \Ha_1)$ such that the map has the form
\[
(\Lambda_1,\dots,\Lambda_m)\mapsto \sum_j \xi_j\circ(id_D\otimes \Lambda_j)\circ\xi
\]
This seems to be more general than the supermaps considered in \cite{daria_super},
more precisely, this map consists of  $m$ supermaps in the sense of \cite{daria_super},  which have the first channel equal to the same $\xi$.

\qed

\end{ex}

The decomposition given in this section can be understood as a physical
realization of generalized supermaps in $\Ce_J(\Be_0,\dots,\Be_n)$. It is not unique,
indeed, for example,  by Theorem \ref{thm:gen_chanels_on_chans}, any state
$\rho$ on $\Ha_0\otimes \Ha_A$ and a POVM   on $B(\Ha_1\otimes \Ha_A)$ define a
PPOVM, but (by the first part of this Theorem), we can always have a
decomposition where the  state is pure. The elements in
$\Ce_J(\Be_0,\dots,\Be_{n+1})$ do not distinguish between these different
realizations, but only the generalized channels they define. We may go a step
further and consider maps which recognize only the channels on $K$, defined by
the generalized channels, that is, maps which give the same result on equivalent
channels. This is the content of the next paragraph.

\subsection{Equivalence of generalized supermaps}

By Theorems \ref{thm:gchannel} and \ref{thm:gchansongchans},
two elements $X_1,X_2\in \Ce_J(\Be_0,\dots,\Be_n)$ are equivalent if and
only if
\begin{equation}\label{eq:equivgch}
X_1-X_2\in \Be_n\otimes (J_{n-1}^T)^\perp
\end{equation}
Using Lemma \ref{lemma:perp_0}, we get
\[
(J_{n-1}^T)^\perp=S_{n-1}^*(S_{n-2}^{-1}(S_{n-3}^*(\dots (L^T)^\perp\dots)))
\]
where $(L^T)^\perp=S_1^*([I_\Ae]^\perp\cap J)$ if $n$ is even and
$(L^T)^\perp=S_1^{-1}((J^T)^\perp)$ if $n$ is odd. From this, we get the following Proposition:

\begin{prop}\label{prop:equivalence} Let $k=\lfloor \frac n 2 \rfloor$.
Two elements $X_1,X_2\in \Ce_J(\Be_0,\dots,\Be_n)$ are equivalent if and only if there are elements
$W^{(m)}\in \Be_n\otimes \Ae_{n-2m}$, $m=1,\dots k$, such
 that
\begin{eqnarray*}
X_1-X_2&=&I_{\Be_{n-1}}\otimes W^{(1)}\\
\ptr_{\Be_{n-2m}}W^{(m)}&=&I_{\Be_{n-2m-1}}\otimes W^{(m+1)},\ m=1,\dots, k-1\\
W^{(k)}&\in& \Be_n\otimes J,\quad \ptr_{\Be_0} W^{(k)}=0\qquad \mbox{if } n=2k\\
\ptr_{\Be_1} W^{(k)}&\in& \Be_n\otimes (J^T)^\perp\qquad \mbox{if } n=2k+1
\end{eqnarray*}

\end{prop}

It is not clear in the present how to interpret this equivalence, in terms of the physical realizations of the channels. The next Theorem gives a characterization
of  elements in
$\Ce_J(\Be_0,\dots,\Be_{n+1})$ which respect this equivalence.

\begin{thm}\label{thm:equiv_gchans} The set of all elements in
$\Ce_J(\Be_0,\dots,\Be_{n+1})$
having the same value on each equivalence   class of elements in $\Ce_J(\Be_0,\dots,\Be_n)$ is
\[
J_{n+1}\cap (\Be_{n+1}\otimes\Be_n\otimes J_{n-1})\cap c_{n+1}\Se (\Ae_{n+1})
\]
In particular, if $K=\Se(\Be_0)$ then this set has the form
\begin{eqnarray*}
\Ce(\Be_0,\dots,\Be_{n+1})&\cap& \Ce(\Be_0,\Be_n,\Be_{n+1},\Be_1,\dots,\Be_{n-1}),\ \mbox{if $n$ is odd}\\
\Ce(\Be_0,\dots,\Be_{n+1})&\cap& \Ce(\Be_n,\Be_{n+1},\Be_0,\dots,\Be_{n-1}),\ \mbox{if $n$ is even}\\
\end{eqnarray*}

\end{thm}

{\it Proof.} Let $X\in \Ce_J(\Be_0,\dots,\Be_{n+1})$, then it is clear from
(\ref{eq:equivgch}) that the corresponding  map has the same value  on equivalent elements if and only if it is equal to 0 on  $\Be_n\otimes (J_{n-1}^T)^\perp$. Equivalently,
\[
0=\Tr \left(b\ptr_{\Ae_n}[(I_{\Be_{n+1}}\otimes Y^T)X]\right)=\Tr ((b\otimes Y^T)X)
\]
for all $b\in \Be_{n+1}$ and $Y\in \Be_n\otimes (J_{n-1}^T)^\perp$, that is,
$X\in (\Be_{n+1}\otimes \Be_n\otimes J_{n-1}^\perp)^\perp=\Be_{n+1}\otimes \Be_n\otimes J_{n-1}$. Since $X\in \Ce_J(\Be_0,\dots,\Be_{n+1})$, we get the result.

Suppose  $K=\Se(\Be_0)$ and let $k=\lfloor \frac{n+1}2\rfloor$.
Since $X\in  \Be_{n+1}\otimes \Be_n\otimes J_{n-1}$,
there are positive elements $Z^{(m)}\in \Be_{n+1}\otimes\Be_n\otimes
\Ae_{n-1-2m}$,  such that
 \begin{eqnarray}
\ptr_{\Be_{n-1-2m}} Z^{(m)}&=&I_{\Be_{n-2-2m}}\otimes Z^{(m+1)},\qquad m=0,\dots,k-2\label{eq:Z1}\\
Z^{(k-1)}&\in& \Be_{n+1}\otimes\Be_n\otimes J \ \mbox{ if $n$ is odd },\label{eq:Zodd}\\
Z^{(k-1)}&\in& \Be_{n+1}\otimes\Be_n\otimes S_1^{-1}(\tilde J)\ \mbox{ if $n$ is even}\label{eq:Zeven}
\end{eqnarray}
and $Z^{(0)}=X$. Suppose $n$ is odd, then by Theorem
\ref{thm:gchans_daria}, we get
\[
\ptr_{\Be_{n+1}}\ptr_{\Be_{n-1}}\dots\ptr_{\Be_{2}} X=I_{\Be_{n}\otimes\Be_{n-2}\otimes\dots\otimes\Be_{1}}\otimes Y^{(k)}
\]
with $Y^{(k)}\in \Se(\Be_0)$, and from (\ref{eq:Z1}), we have
\[
\ptr_{\Be_{n-1}}\ptr_{\Be_{n-3}}\dots\ptr_{\Be_{2}} X=I_{\Be_{n-2}\otimes\Be_{n-4}\otimes\dots\otimes\Be_{1}}\otimes Z^{(k-1)}
\]
This implies $\ptr_{\Be_{n+1}}Z^{(k-1)}=I_{\Be_n}\otimes Y^{(k)}$. If $J=\Be_0$,
this together with (\ref{eq:Z1})
and (\ref{eq:Zodd}) is equivalent with $X\in \Ce(\Be_0, \Be_n,\Be_{n+1},\Be_1,\dots,\Be_{n-1})$.
Similarly, if $J=\Be_0$ and $n$ is even, we have
\[
\ptr_{\Be_{n+1}}\ptr_{\Be_{n-1}}\dots\ptr_{\Be_{1}} X=I_{\Be_{n}\otimes\Be_{n-2}\otimes\dots\otimes\Be_{0}}
\]
and by (\ref{eq:Zeven}), there is some positive element
$Z^{(k)}\in \Be_{n+1}\otimes \Be_n$, such that
\begin{equation}\label{eq:ZZeven}
\ptr_{\Be_1} Z^{(k-1)}=I_{\Be_0}\otimes Z^{(k)}
\end{equation}
Then
\[
\ptr_{\Be_{n-1}}\ptr_{\Be_{n-3}}\dots\ptr_{\Be_{1}} X=I_{\Be_{n-2}\otimes\Be_{n-4}\otimes\dots\otimes\Be_{0}}\otimes Z^{(k)}
\]
so that  we  must have $\ptr_{\Be_{n+1}}Z^{(k)}=I_{\Be_n}$. This, together with (\ref{eq:Z1}) and (\ref{eq:ZZeven}), is equivalent
 with $X\in \Ce(\Be_n,\Be_{n+1},\Be_0,\dots,\Be_{n-1})$.

\qed

\begin{ex}(\textbf{Equivalence on PPOVMs})
Suppose  that $Z$ is a generalized POVM on the set of PPOVMs,
that is, $Z\in \Ce(B(\Ha_0),B(\Ha_1),\mathbb C^m,\mathbb C^k)$. Then by Example \ref{ex:chan_on_PPOVM},
$Z=\sum_{i=1}^k\sum_{j=1}^m |i\>\<i|\otimes |j\>\<j|\otimes Z_{ij}$ and
each $Z_{ij}$ is the Choi matrix of a cp map $\Lambda_{ij}:B(\Ha_0)\to B(\Ha_1)$,
such that there is a channel $\xi$ with
$\sum_j\Lambda_{ij}=\xi$ for all $i$.
If $Z$ attains the same value on
equivalent elements, then it defines a measurement on the set of equivalence classes of PPOVMs, that is,
 on the set of measurements on channels $B(\Ha_0)\to B(\Ha_1)$. By Theorem \ref{thm:equiv_gchans}, this happens
 if and only if $Z$ is also in $\Ce(\mathbb C^m,\mathbb C^k,B(\Ha_0),B(\Ha_1))$. Using Theorem \ref{thm:gchans_daria},
 we get that there are some numbers $\mu_{ij}\ge 0$, with $\sum_j\mu_{ij}=1$ for all $i$, such that
 $\ptr_{\Ha_1} Z_{ij}=\mu_{ij} I_{\Ha_0}$. It follows that there are channels $\xi_{ij}$, such that
 $\Lambda_{ij}=\mu_{ij}\xi_{ij}$. We have proved the following:

 For any  measurement on measurements on $\Ce(B(\Ha_0),B(\Ha_1))$ with values in $\{1,\dots,m\}$, there
are  $\xi_{ij}\in \Ce(B(\Ha_0),B(\Ha_1))$ and numbers $\mu_{ij}\ge 0$, $\sum_{j}\mu_{ij}=1$, satisfying
$\sum_j \mu_{ij}\xi_{ij}=\xi$ for all $i$, such that, if a measurement on
$\Ce(B(\Ha_0),B(\Ha_1))$ has an implementation $(\Ha_D,(M_1,\dots,M_m),\rho)$, then the corresponding probabilities are given by
\[
p_i(\Ha_D,(M_1,\dots,M_m),\rho)=\sum_j\mu_{ij}\Tr( M_j(\xi_{ij}\otimes id_D)(\rho))
\]
Conversely, any such $\xi_{ij}$, $\mu_{ij}$ define a measurement on measurements on \linebreak$\Ce(B(\Ha_0),B(\Ha_1))$. Note that if $(\Ha_D,M,\rho)$ and $(\Ha_E,N,\sigma)$ are implementations of PPOVMs, then these are equivalent if and only if
$\Tr( M_j(\xi\otimes id_D)(\rho)=\Tr( N_j(\xi\otimes id_E)(\sigma)$ for any
channel $\xi$.

\qed
\end{ex}

\subsection{Equivalence of  combs}

Any $N$-comb $X\in \comb(\Be_0,\dots,\Be_{2N+1})$ is a cp map $\comb(\Be_1,\dots,\Be_{2N})\to
\Be_{2N+1}\otimes\Be_0$. By  (\ref{eq:combs_subspace}) and Theorem \ref{thm:gchannel}, two $N$-combs $X_1$ and $X_2$ are equivalent if and only if
\[
X_1-X_2\in \Be_{2N+1}\otimes (\hat J_{2N-1}^T)^\perp\otimes \Be_0
\]
where $\hat J_{2N-1}:=J_{2N-1}(\Be_1,\dots,\Be_{2N})$.

\begin{prop} Two elements $X_1,X_2\in \comb(\Be_0,\dots,\Be_{2N-1})$ are equivalent
if and only if there are elements $V^{(m)}\in \Be_{2N+1}\otimes \Ae_{2m-1}$, $m=1,\dots,N$,
such that
\begin{eqnarray*}
X_1-X_2&=&I_{\Be_{2N}}\otimes V^{(N)}\\
\ptr_{\Be_{2m-1}} V^{(m)}&=&I_{\Be_{2m-2}}\otimes V^{(m-1)},\quad m=2,\dots,N\\
\ptr_{\Be_1} V^{(1)}&=&0
\end{eqnarray*}
\end{prop}

The proof of the next Theorem is the same as of Theorem \ref{thm:equiv_gchans}.
\begin{thm}
The set elements in  $\comb(\Be_0,\dots,\Be_{2N+1})$ having the same value
on equivalent elements in $\comb(\Be_1,\dots,\Be_{2N})$ is equal to
\[
\comb(\Be_0,\dots,\Be_{2N+1})\cap \comb(\Be_0,\Be_1,\Be_{2N},\Be_{2N+1},\Be_2,\dots,\Be_{2N-1})
\]

\end{thm}

\section{Final remarks}

We  have introduced the concept of a channel on a section
 of the state space of a finite dimensional $C^*$-algebra. We proved that such
channels are
restrictions of completely positive maps, called generalized channels. If the section $K$ contains the tracial state, the Choi matrices of generalized channels with respect to $K$ form
again a section of the state space of some $C^*$-algebra. This allows us to define generalized supermaps as completely positive maps sending generalized channels
(or generalized supermaps) to states.
The set of generalized supermaps is characterized as an intersection of
the state space by a subspace. This  might be useful, for example,
 in optimization  problems with respect to supermaps.

Although the condition $\tau_\Ae\in K$ includes the most important examples of
channels and combs, it might be interesting to consider supermaps for arbitrary
generalized channels. By Proposition \ref{prop:set_of_gchans}, this
 should be possible by extending
our  theory using the set $\Se_\rho(\Ae)$ instead of $\Se(\Ae)$, with
 an invertible element $\rho\in \Ae^+$. This can be done along similar lines.

Another possible extension of the theory  is to look at the generalized
channels sending  a section $K_1$ to a given convex subset $K_2$ of the target
state  space.
The set $\comb_J(\Ae,\Be,\Ce)$ is a particular example of this, but arbitrary
 convex subset can be considered, using similar tools as were used in the present paper.

A natural question is an extension of these results to infinite dimension. For
example,  in the setting of the algebras of bounded operators $B(\Ha)$ for infinite dimensional Hilbert space $\Ha$, quantum supermaps were studied in 
\cite{umanita}. Channels and measurements on sections of the state space can be 
studied also in this case and similar results can be expected. But the identification of the set of channels with a section of a state space fails.

\section*{Acknowledgements}

I would like to thank M. Ziman and T. Heinosaari for useful comments and 
discussions. I would
also like to thank the anonymous referee for a number of valuable comments
and suggestions,  especially for Theorem \ref{thm:arveson} and its proof,
which greatly improved the paper.  

The work was supported by the grants 
VEGA 2/0032/09 and meta-QUTE ITMS 26240120022.

\section*{Appendix}

Let $\Ae=\oplus_n B(\Ha_{A_n})$ be a finite dimensional $C^*$ algebra and let
$\Ha_A$, $\Ha_B$, $\Ha_B'$ be finite dimensional Hilbert spaces. Let
$ T: \Ae\otimes B(\Ha_B)\to B(\Ha_{AB'})$ be a cp map. Then we say that $T$
is semicausal if
\begin{equation}\label{eq:semicau}
T(I_\Ae\otimes b)=I_A\otimes S(b)
\end{equation}
for some cp map $S: B(\Ha_B)\to B(\Ha_{B'})$, and $T$ is semilocalizable, if
\begin{equation}\label{eq:semiloc}
T=(id_A\otimes G)\circ(F\otimes id_B)
\end{equation}
for some  unital cp map $F: \Ae\to B(\Ha_{AD})$ and a cp map
$G: B(\Ha_{DB})\to B(\Ha_{B'})$, where $\Ha_D$ is some (finite dimensional)
Hilbert space. The following statement was proved in \cite{esw},
 in the case that $\Ae$ is a matrix algebra. For the convenience of the reader, we give the modification of the proof in \cite{esw} for our slightly more general case.

\begin{lemma}\label{lemma:apndx} Let
$T: \Ae\otimes B(\Ha_B)\to \Ha_{AB'}$ be a cp map. Then $T$ is semicausal if and
 only if $T$ is semilocalizable.

\end{lemma}

{\it Proof.} Any representation  of
$\Ae\otimes B(\Ha_B)$ has the form
 \[
\Pi(a\otimes b)= \oplus_n I_{E_n}\otimes a_n\otimes b = (\oplus_n I_{E_n}\otimes a_n)\otimes b
\]
for some Hilbert spaces $\Ha_{E_n}$, where  $a=\oplus_n a_n\in \Ae$ and
$b\in B(\Ha_B)$. Hence
by Stinespring representation,  $T$ has the form
\[
T(a\otimes b)=V^*((\oplus_n I_{E_n}\otimes a_n)\otimes b)V
\]
for some linear map $V: \Ha_{AB'}\to \oplus_n \Ha_{E_nA_nB}$. Let now
\[
S(b)=W^*(1_D\otimes b)W
\]
be a minimal Stinespring representation of $S$. Then (\ref{eq:semicau}) implies
that
\[
V^*(I_{\oplus_n\Ha_{E_nA_n}}\otimes b)V= (I_A\otimes W^*)(I_{AD}\otimes b)(W\otimes I_B)
\]
Exactly as in \cite{esw}, we get by minimality of the Stinespring representation that there is
some isometry $U: \Ha_{AD}\to \oplus_n \Ha_{E_nA_n}$, such that
\[
V=(U\otimes I_B)(I_A\otimes W)
\]
Hence
\[
\Phi(a\otimes b)= (I_A\otimes W^*)(U^*(\oplus_n I_{E_n}\otimes a_n)U\otimes b )(I_A\otimes W)
\]
so that
\begin{equation}\label{eq:semiloc1}
\Phi=(id_A\otimes G)\circ(F\otimes id_B)
\end{equation}
for the unital cp map $F: \Ae\to B(\Ha_{AD})$, given by $F(a)=
U^*(\oplus_n I_{E_n}\otimes a_n)U$ and the cp map
$G: B(\Ha_{DB})\to B(\Ha_{B'})$, defined as $G(d\otimes b)=W^*(d\otimes b)W$.

Conversely, if $T$ is of the form (\ref{eq:semiloc1}), then it is clear that
$T$ satisfies (\ref{eq:semicau}), with
\begin{equation}\label{eq:xi}
S(b)=G(1_D\otimes b)
\end{equation}

\qed

\begin{thm}\label{thm:apndx} Let $\Ae=\oplus B(\Ha_{A_k})$, $\Be=\oplus B(\Ha_{B_m})$, $\Ce=\oplus B(\Ha_{C_n})$ be
finite dimensional $C^*$ algebras, with
minimal central projections $\{p_k\}_k$, $\{q_m\}_m$ and $\{r_n\}_n$,
respectively. Let $X\in \Ae\otimes\Be\otimes\Ce$ be positive. Then the following are equivalent.
\begin{enumerate}
\item[(i)] There is some positive element  $Y\in \Ce$ such  that
\[
\ptr_\Ae X=I_\Be\otimes Y
\]
\item[(ii)] There is an auxiliary  Hilbert space $\Ha_{D}$,  positive elements
$X_0(n)\in B(\Ha_{DC_n})$ and $X_1(m,n)\in \Ce(B(\Ha_{B_mD}),\Ae)$
such that
\[
X_{m,n}:=(I_\Ae\otimes q_m\otimes r_n)X= X_1(m,n)*X_0(n)
\]
\end{enumerate}
Moreover, we have
\[
\ptr_{D}X_0(n)=Y_n:=r_nY
\]
\end{thm}

{\it Proof.} Suppose first that $\Be=B(\Ha_B)$ and $\Ce=B(\Ha_C)$ are matrix algebras.
 We can always write  $\Ha_C=\Ha_{C_1}\otimes \Ha_{C_2}$. Let us define the map
$\Phi: B(\Ha_{BC_1})\to \Ae\otimes B(\Ha_{C_2})$ by
\[
\Phi(a)=X*a,\qquad a\in  B(\Ha_{BC_1})
\]
 Then $\Phi$ is a cp map and
\[
\ptr_{\Ae}\Phi(a)=[\ptr_\Ae X]*a,\qquad a\in  B(\Ha_{BC_1})
\]
so that $\ptr_\Ae X$ is the Choi matrix of $\ptr_\Ae\circ \Phi$.
Similarly, if $\xi:B(\Ha_{C_1})\to B(\Ha_{C_2})$ is the cp map with C-J matrix $Y$, then $I_\Ae\otimes Y$ is the C-J matrix of $\xi\circ \ptr_\Ae$.
It follows that the maps $\Phi$ and $\xi$ satisfy
\[
\ptr_\Ae\circ \Phi=\xi\circ \ptr_\Ae
\]
For the adjoints, this condition has he form
$\Phi^*(I_\Ae\otimes c)=I_B\otimes \xi^*(c)$, for all $ c\in B(\Ha_{C_2})$
which means that the map $\Phi^*$ is semicausal. By  Lemma \ref{lemma:apndx},
 (i) is equivalent with
\[
\Phi=(F^*\otimes id_{C_2})\circ (id_B\otimes G^*)
\]
for a cp map  $G^*:B(\Ha_{C_1})\to B(\Ha_{DC_2})$ and a channel
 $F^*:B(\Ha_{BD})\to \Ae$, with some Hilbert space $\Ha_D$.
By putting $X_1$ and $X_0$ the Choi matrices of $F$ and $G$, respectively,
we get (ii).  Finally, (\ref{eq:xi}) implies  $\ptr_D X_0=Y$.

For the general case, note that  $X_{m,n}\in \Ae\otimes B(\Ha_{B_mC_n})$ and
\[
\ptr_\Ae X_{m,n}=(q_m\otimes r_n)\ptr_\Ae X,
\]
so that (i) is equivalent with
\[
\ptr_\Ae X_{m,n}=I_{B_m}\otimes Y_n,\qquad \forall m,n
\]
where $Y_n=r_n Y\in B(\Ha_{C_n})^+$. By the first part of the proof,
we get that (i) holds if and only if
\[
X_{m,n}=X'_1(m,n)*X'_0(m,n)
\]
with positive elements
$X'_0(m,n)\in B(\Ha_{D_{m,n}C_n})$, $X'_1(m,n)\in \Ce(B(\Ha_{B_mD_{m,n}}),\Ae)$
 for some
ancillary Hilbert spaces $\Ha_{D_{m,n}}$, and such that
$\ptr_{D_{m,n}} X'_0(m,n)=Y_n$.
 Note further that in the proof of Lemma \ref{lemma:apndx}, the cp map $G$ and
the ancilla $\Ha_D$ are
given by a minimal Stinespring representation of $S$. Hence $X'_0(m,n)$ and the
 ancilla are determined by $Y_n$, so that these depend only from $n$.
Moreover, there are some $\Ha_{D'_n}$ and $\Ha_D$, such that $\Ha_D=\Ha_{D_nD'_n}$ for all $n$. Choose some state $\omega_n\in B(\Ha_{D'_n})$ for all $n$ and put
\[
X_0(n):=\omega_n\otimes X_0'(n),\quad    X_1(m,n):=X_1'(m,n)\otimes I_{D_n'}
\]
Then $X_0(n)\in B(\Ha_{DC_n})$, $ X_1(m,n)\in \Ce(B(\Ha_{B_mD}),\Ae)$ and
\[
X_1(m,n)*X_0(n)=X_1(m,n)*I_{D_n'}*\omega_n*X'_0(n)=X_1'(m,n)*X_0'(n)=X_{m,n}
\]
Clearly also
\[
\ptr_DX_0(n)=\ptr_{D_n}X_0'(n)=Y_n.
\]

\qed

\end{document}